\documentclass[preprint,12pt]{elsarticle}

\newcommand\modif[1]
{\textcolor{black}{#1}}
\newcommand\kendall[1]
{\textcolor{black}{#1}}

\usepackage{amssymb}
\usepackage{ulem}
\usepackage[usenames]{color}

\journal{Elsevier}
\usepackage{ifpdf}
\ifpdf

\begin{document}

\begin{frontmatter}



\title{
Measurements of the T2K neutrino beam \modif{properties} using the INGRID on-axis near detector
}

\newcommand\revaffiliation[1]{#1}
\newcommand\revthanks[1]{\thanks{#1}}
\newcommand\revauthor[2]{\author[#2]{#1}}

\revauthor{K.\,Abe}{INSTBJ}
\revauthor{N.\,Abgrall}{INSTEG}
\revauthor{Y.\,Ajima\revthanks{also at J-PARC Center}}{INSTCB}
\revauthor{H.\,Aihara}{INSTCH}
\revauthor{J.B.\,Albert}{INSTFH}
\revauthor{C.\,Andreopoulos}{INSTEH}
\revauthor{B.\,Andrieu}{INSTBB}
\revauthor{M.D.\,Anerella}{INSTBR}
\revauthor{S.\,Aoki}{INSTCC}
\revauthor{O.\,Araoka\revthanks{also at J-PARC Center}}{INSTCB}
\revauthor{J.\,Argyriades}{INSTEG}
\revauthor{A.\,Ariga}{INSTEE}
\revauthor{T.\,Ariga}{INSTEE}
\revauthor{S.\,Assylbekov}{INSTFG}
\revauthor{D.\,Autiero}{INSTJ}
\revauthor{A.\,Badertscher}{INSTEF}
\revauthor{M.\,Barbi}{INSTE}
\revauthor{G.J.\,Barker}{INSTFD}
\revauthor{G.\,Barr}{INSTGG}
\revauthor{M.\,Bass}{INSTFG}
\revauthor{M.\,Batkiewicz}{INSTDG}
\revauthor{F.\,Bay}{INSTEE}
\revauthor{S.\,Bentham}{INSTEJ}
\revauthor{V.\,Berardi}{INSTGF}
\revauthor{B.E.\,Berger}{INSTFG}
\revauthor{I.\,Bertram}{INSTEJ}
\revauthor{M.\,Besnier}{INSTBA}
\revauthor{J.\,Beucher}{INSTI}
\revauthor{D.\,Beznosko}{INSTFJ}
\revauthor{S.\,Bhadra}{INSTH}
% ...

\revauthor{\\(The T2K Collaboration)}{}

\cortext[cor1]{Corresponding author.}

\newcommand\revaddress[2]{\address[#1]{#2}}

\revaddress{INSTC}{\revaffiliation{University of Alberta, Centre for Particle Physics, Department of Physics, Edmonton, Alberta, Canada}}
\revaddress{INSTDF}{\revaffiliation{National Center for Nuclear Research, Warsaw, Poland}}
\revaddress{INSTEE}{\revaffiliation{University of Bern, Albert Einstein Center for Fundamental Physics, Laboratory for High Energy Physics (LHEP), Bern, Switzerland}}
\revaddress{INSTFE}{\revaffiliation{Boston University, Department of Physics, Boston, Massachusetts, U.S.A.}}
\revaddress{INSTD}{\revaffiliation{University of British Columbia, Department of Physics and Astronomy, Vancouver, British Columbia, Canada}}
\revaddress{INSTBR}{\revaffiliation{Brookhaven National Laboratory, Physics Department, Upton, New York, U.S.A.}}
\revaddress{INSTGA}{\revaffiliation{University of California, Irvine, Department of Physics and Astronomy, Irvine, California, U.S.A.}}
\revaddress{INSTI}{\revaffiliation{IRFU, CEA Saclay, Gif-sur-Yvette, France}}
\revaddress{INSTCI}{\revaffiliation{Chonnam National University, Institute for Universe \& Elementary Particles, Gwangju, Korea}}
\revaddress{INSTGB}{\revaffiliation{University of Colorado at Boulder, Department of Physics, Boulder, Colorado, U.S.A.}}
\revaddress{INSTFG}{\revaffiliation{Colorado State University, Department of Physics, Fort Collins, Colorado, U.S.A.}}
\revaddress{INSTCJ}{\revaffiliation{Dongshin University, Department of Physics, Naju, Korea}}
\revaddress{INSTFH}{\revaffiliation{Duke University, Department of Physics, Durham, North Carolina, U.S.A.}}
\revaddress{INSTBA}{\revaffiliation{Ecole Polytechnique, IN2P3-CNRS, Laboratoire Leprince-Ringuet, Palaiseau, France }}
\revaddress{INSTEF}{\revaffiliation{ETH Zurich, Institute for Particle Physics, Zurich, Switzerland}}
\revaddress{INSTEG}{\revaffiliation{University of Geneva, Section de Physique, DPNC, Geneva, Switzerland}}
\revaddress{INSTDG}{\revaffiliation{H. Niewodniczanski Institute of Nuclear Physics PAN, Cracow, Poland}}
\revaddress{INSTCB}{\revaffiliation{High Energy Accelerator Research Organization (KEK), Tsukuba, Ibaraki, Japan}}
\revaddress{INSTED}{\revaffiliation{Institut de Fisica d'Altes Energies (IFAE), Bellaterra (Barcelona), Spain}}
\revaddress{INSTEC}{\revaffiliation{IFIC (CSIC \& University of Valencia), Valencia, Spain}}
\revaddress{INSTEI}{\revaffiliation{Imperial College London, Department of Physics, London, United Kingdom}}
\revaddress{INSTGF}{\revaffiliation{INFN Sezione di Bari and Universit\`a e Politecnico di Bari, Dipartimento Interuniversitario di Fisica, Bari, Italy}}
\revaddress{INSTNA}{\revaffiliation{INFN Sezione di Napoli and Universit\`a di Napoli, Dipartimento di Fisica, Napoli, Italy}}
\revaddress{INSTPA}{\revaffiliation{INFN Sezione di Padova and Universit\`a di Padova, Dipartimento di Fisica, Padova, Italy}}
\revaddress{INSTBD}{\revaffiliation{INFN Sezione di Roma and Universit\`a di Roma "La Sapienza", Roma, Italy}}
\revaddress{INSTEB}{\revaffiliation{Institute for Nuclear Research of the Russian Academy of Sciences, Moscow, Russia}}
\revaddress{INSTCC}{\revaffiliation{Kobe University, Kobe, Japan}}
\revaddress{INSTCD}{\revaffiliation{Kyoto University, Department of Physics, Kyoto, Japan}}
\revaddress{INSTEJ}{\revaffiliation{Lancaster University, Physics Department, Lancaster, United Kingdom}}
\revaddress{INSTFC}{\revaffiliation{University of Liverpool, Department of Physics, Liverpool, United Kingdom}}
\revaddress{INSTFI}{\revaffiliation{Louisiana State University, Department of Physics and Astronomy, Baton Rouge, Louisiana, U.S.A.}}
\revaddress{INSTJ}{\revaffiliation{Universit\'e de Lyon, Universit\'e Claude Bernard Lyon 1, IPN Lyon (IN2P3), Villeurbanne, France}}
\revaddress{INSTCE}{\revaffiliation{Miyagi University of Education, Department of Physics, Sendai, Japan}}
\revaddress{INSTFJ}{\revaffiliation{State University of New York at Stony Brook, Department of Physics and Astronomy, Stony Brook, New York, U.S.A.}}
\revaddress{INSTCF}{\revaffiliation{Osaka City University, Department of Physics, Osaka,  Japan}}
\revaddress{INSTGG}{\revaffiliation{Oxford University, Department of Physics, Oxford, United Kingdom}}
\revaddress{INSTBB}{\revaffiliation{UPMC, Universit\'e Paris Diderot, CNRS/IN2P3, Laboratoire de Physique Nucl\'eaire et de Hautes Energies (LPNHE), Paris, France}}
\revaddress{INSTGC}{\revaffiliation{University of Pittsburgh, Department of Physics and Astronomy, Pittsburgh, Pennsylvania, U.S.A.}}
\revaddress{INSTFA}{\revaffiliation{Queen Mary, University of London, School of Physics and Astronomy, London, United Kingdom}}
\revaddress{INSTE}{\revaffiliation{University of Regina, Physics Department, Regina, Saskatchewan, Canada}}
\revaddress{INSTGD}{\revaffiliation{University of Rochester, Department of Physics and Astronomy, Rochester, New York, U.S.A.}}
\revaddress{INSTBC}{\revaffiliation{RWTH Aachen University, III. Physikalisches Institut, Aachen, Germany}}
\revaddress{INSTDD}{\revaffiliation{Seoul National University, Department of Physics and Astronomy, Seoul, Korea}}
\revaddress{INSTFB}{\revaffiliation{University of Sheffield, Department of Physics and Astronomy, Sheffield, United Kingdom}}
\revaddress{INSTDI}{\revaffiliation{University of Silesia, Institute of Physics, Katowice, Poland}}
\revaddress{INSTDA}{\revaffiliation{STFC, Daresbury Laboratory, Warrington, United Kingdom}}
\revaddress{INSTEH}{\revaffiliation{STFC, Rutherford Appleton Laboratory, Harwell Oxford, United Kingdom}}
\revaddress{INSTCH}{\revaffiliation{University of Tokyo, Department of Physics, Tokyo, Japan}}
\revaddress{INSTBJ}{\revaffiliation{University of Tokyo, Institute for Cosmic Ray Research, Kamioka Observatory, Kamioka, Japan}}
\revaddress{INSTCG}{\revaffiliation{University of Tokyo, Institute for Cosmic Ray Research, Research Center for Cosmic Neutrinos, Kashiwa, Japan}}
\revaddress{INSTF}{\revaffiliation{University of Toronto, Department of Physics, Toronto, Ontario, Canada}}
\revaddress{INSTB}{\revaffiliation{TRIUMF, Vancouver, British Columbia, Canada}}
\revaddress{INSTG}{\revaffiliation{University of Victoria, Department of Physics and Astronomy, Victoria, British Columbia, Canada}}
\revaddress{INSTDJ}{\revaffiliation{University of Warsaw, Faculty of Physics, Warsaw, Poland}}
\revaddress{INSTDH}{\revaffiliation{Warsaw University of Technology, Institute of Radioelectronics, Warsaw, Poland}}
\revaddress{INSTFD}{\revaffiliation{University of Warwick, Department of Physics, Coventry, United Kingdom}}
\revaddress{INSTGE}{\revaffiliation{University of Washington, Department of Physics, Seattle, Washington, U.S.A.}}
\revaddress{INSTGH}{\revaffiliation{University of Winnipeg, Department of Physics, Winnipeg, Manitoba, Canada}}
\revaddress{INSTEA}{\revaffiliation{Wroclaw University, Faculty of Physics and Astronomy, Wroclaw, Poland}}
\revaddress{INSTH}{\revaffiliation{York University, Department of Physics and Astronomy, Toronto, Ontario, Canada}}

\begin{abstract}
Precise measurement of neutrino beam direction and intensity was achieved based on a new concept with modularized neutrino detectors. 
INGRID (Interactive Neutrino GRID) is an on-axis near detector for the T2K long baseline neutrino oscillation experiment. 
INGRID consists of 16 identical modules arranged in horizontal and vertical arrays around the beam center. 
The module has a sandwich structure of iron target plates and scintillator trackers. 
INGRID directly monitors the muon neutrino beam profile center and intensity using the number of observed neutrino events in each module. 
The neutrino beam direction is measured with accuracy better than 0.4~mrad from the measured profile center. 
The normalized event rate is measured with 4\% precision. 
\end{abstract}

\begin{keyword}
Neutrino oscillation \sep T2K \sep Neutrino beam \sep Neutrino detector \sep extruded scintillator \sep wave length shifting fiber
\end{keyword}

\end{frontmatter}

\section{Introduction}\label{sec:intro}
To investigate the flavor mixing of neutrinos\modif{~\cite{bib:PMNS_matrix, bib:PMNS_matrix2}} and the neutrino mass splittings, long baseline (over 100~km) neutrino oscillation experiments are currently running and also being prepared. 
In these experiments, it is extremely important to measure the beam direction and intensity to ensure the stable neutrino production from a primary proton beam. 
\par
\begin{figure}[htbp]
 \begin{center}
  \includegraphics[width=0.75\textwidth]
	{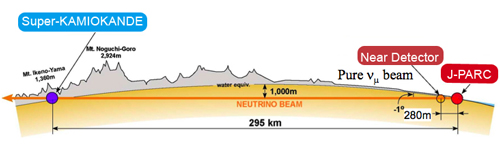}
  \caption{\kendall{Schematic view of} the T2K experiment configuration. At the near detector location, there are two detectors: one is an off-axis near detector and the other is the INGRID on-axis near detector.}
  \label{fig:T2Koverview}
 \end{center}
\end{figure}

T2K (Tokai-to-Kamioka)~\cite{bib:T2KNIM} is a long baseline neutrino oscillation experiment. 
An intense muon neutrino beam is produced by using the 30-GeV proton synchrotron at J-PARC (Japan Proton Accelerator Research Complex) in Tokai. 
The proton beam impinges on a graphite target to produce charged pions, which are focused by three magnetic horns. 
The pions decay mainly into muon -- muon-neutrino pairs during their passage through the 96-meter decay volume. 
After traveling 295 km, the neutrinos are detected by the Super-Kamiokande (SK) detector~\cite{bib:SK} in the Kamioka Observatory (Fig.~\ref{fig:T2Koverview}). 
The goals of \kendall{the T2K experiment} are to measure oscillation parameters with \kendall{a} precision of $\delta(\sin^{2}2\theta_{23})\sim0.01$ and $\delta(\Delta m^2_{23})\sim10^{-4}\ \mathrm{eV^{2}}$ via $\nu_{\mu}$ disappearance and to investigate $\theta_{13}$ through $\nu_{\mu}\rightarrow\nu_{e}$ oscillation. 
\par
T2K adopts an off-axis beam configuration~\cite{bib:offaxisbeam}; the beam center direction is 2.5 degrees away from the direction of SK so that the muon neutrino beam has a narrow energy peak at $\sim0.6$~GeV, which maximizes the effect of the neutrino oscillation at SK and minimizes the background for the signal. 
The neutrino energy, however, varies as a function of the off-axis angle. 
Therefore, it is important to monitor and control the beam direction precisely; 
the beam direction is required to be controlled within \modif{$\pm1$~mrad}. 
In addition, monitoring of the beam intensity is important to ensure stable beam neutrino production. 
\modif{T2K has a mumon monitor~\cite{bib:MUMON} downstream of the beam dump. 
It measures the beam direction and stability by detecting muons from pion decay for every bunch. 
Since the muon monitor detects only high energy muons which penetrate the beam dump, the phase space of parent pions covered by the muon monitor is much different from the one of pions which produce neutrinos to the near or far detectors. 
INGRID (Interactive Neutrino GRID) is an on-axis near detector and measures the neutrino beam direction and intensity by detecting neutrino interaction events. 
The covered phase space of the parent pions are much closer to the one for the off-axis neutrino detectors than the muon monitor. 
Therefore, the measurement by INGRID is more directly connected with the T2K neutrino property. 
It was designed to provide daily measurements at the design beam intensity (750 kW primary proton beam power). }
\par
This paper reports neutrino beam measurements using the INGRID detector based on the first two physics runs: 
Run 1 (Jan.-Jun. 2010) and Run 2 (Nov. 2010 - Mar. 2011). 
During this time period, INGRID recorded more than 99.6\% of delivered beam corresponding to $1.44\times10^{20}$ protons on target (POT). 
Using data from the same period, the T2K experiment observed indications of $\nu_{\mu}\rightarrow\nu_{e}$ appearance~\cite{bib:T2Kfirstnue}. 
\par
In section 2 and 3, the design and basic performance of the INGRID detector are described, respectively. 
Details of Monte Carlo (MC) simulations and criteria for neutrino event selection are described in section 4 and 5, respectively. 
In section 6, results of neutrino beam measurements with INGRID are summarized. 

\section{Detector configuration}\label{sec:ingrid}
\begin{figure}[htbp]
 \begin{center}
  \includegraphics[width=0.75\textwidth]
	{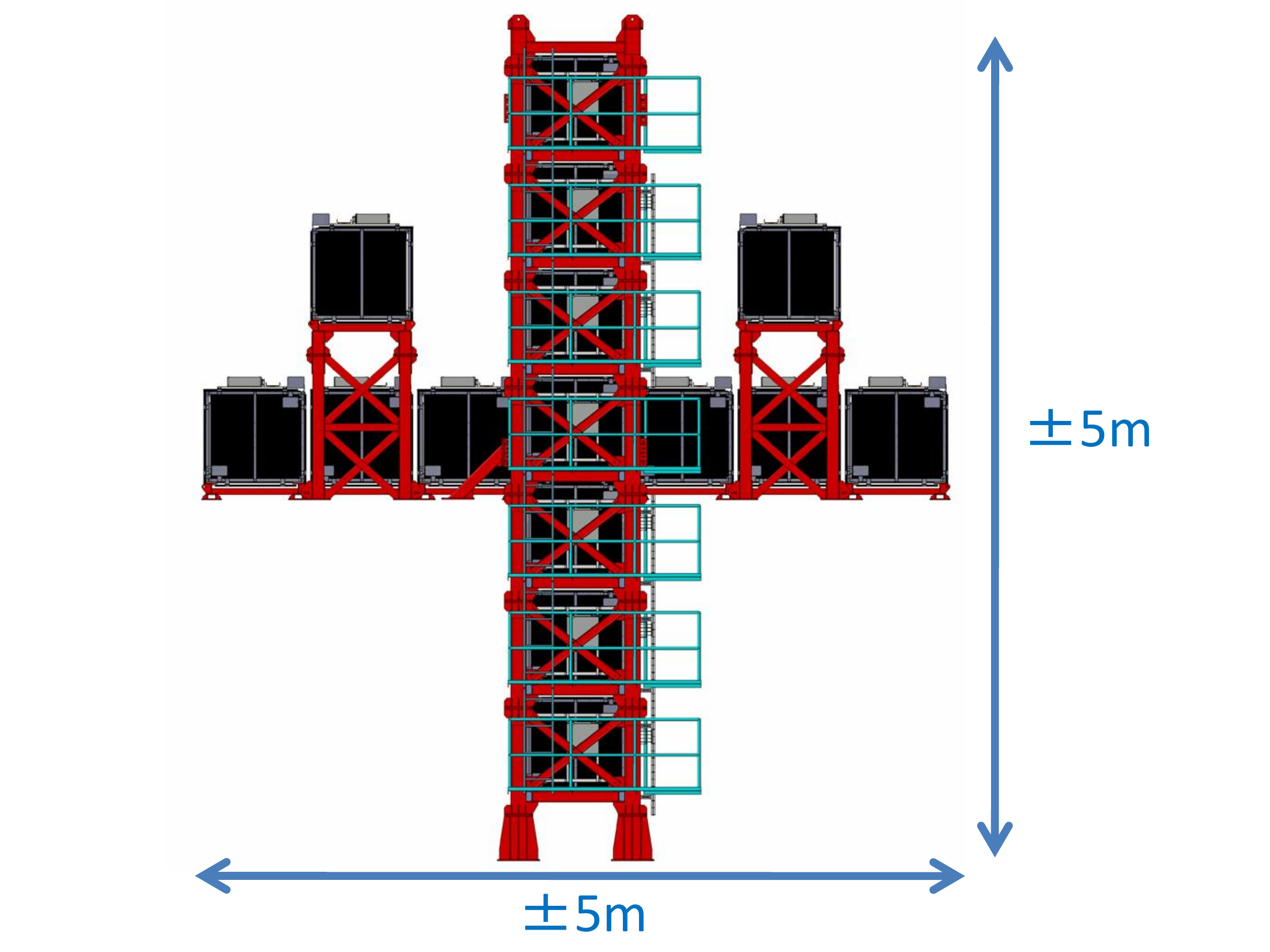}
  \caption{INGRID on-axis near detector. The 16 identical modules sample the neutrino beam in a transverse section of 10 m $\times$ 10 m . The center of the cross, with two overlapping modules, corresponds to the designed neutrino beam center ($\theta=0^{\circ}$).}
  \label{fig:INGRID_overview}
 \end{center}
\end{figure}

INGRID is located 280~meters downstream of the meson production target, where the spatial width (1$\sigma$) of the neutrino beam is about 5~meters.
Therefore, it is designed to sample the beam in a transverse section of 10~m $\times$ 10~m with 14 identical modules arranged as two identical groups along the horizontal and vertical axis. 
Two separate modules are placed at off-axis positions off the main cross, as shown in Fig.~\ref{fig:INGRID_overview} to monitor the asymmetry of the beam. 
\par
\begin{figure}[htbp]
 \begin{center}
  \includegraphics[width=0.75\textwidth]
	{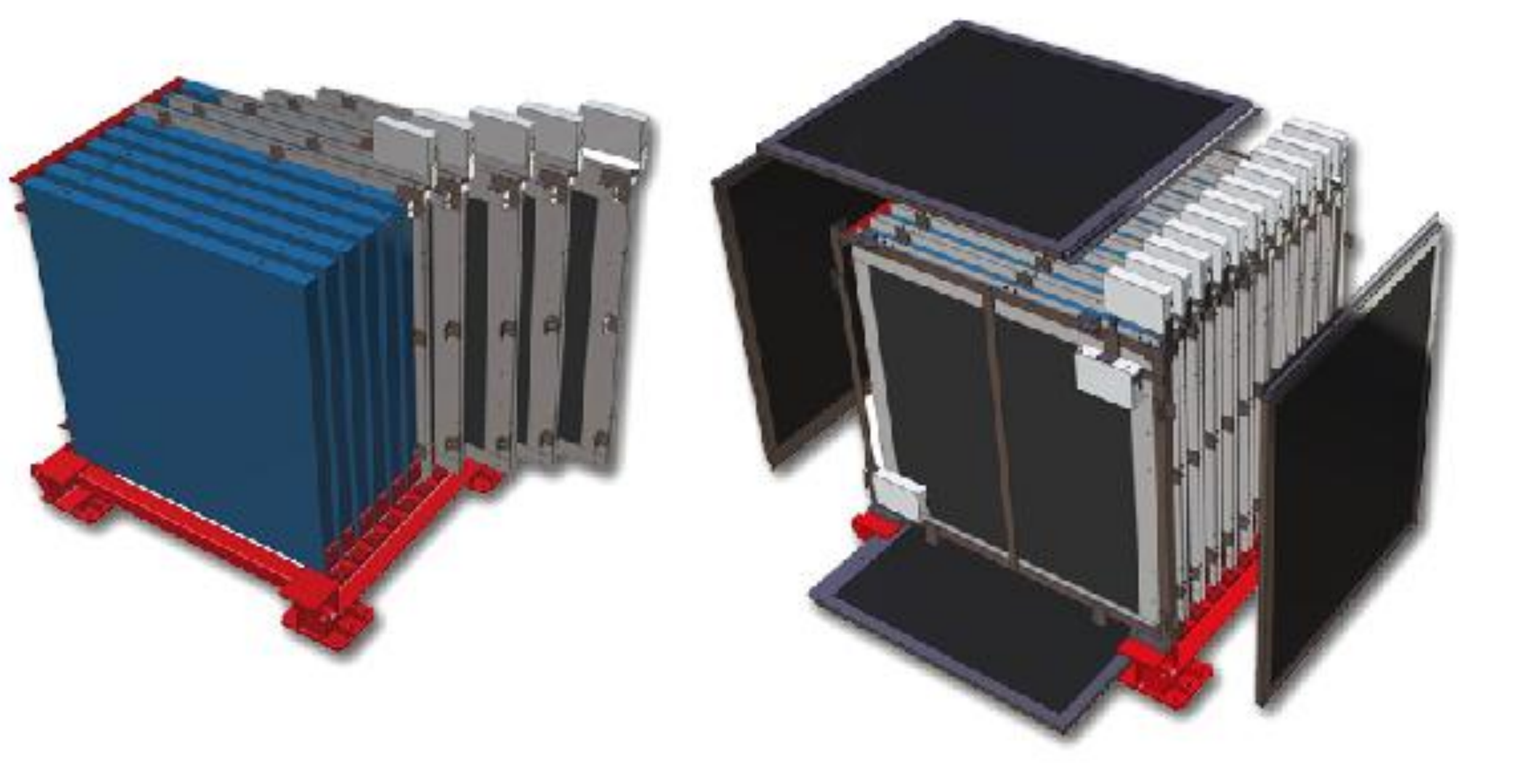}
  \caption{Structure of the module. It is a sandwich made of nine iron target plates and eleven scintillator trackers (left). 
The module is surrounded by the scintillator veto planes (right). }
  \label{fig:INGRID_module}
 \end{center}
\end{figure}

Each of the \kendall{modules} consists of a sandwich structure of nine iron target plates and 11 tracking scintillator planes as shown in Fig.~\ref{fig:INGRID_module} left. 
They are surrounded by veto scintillator planes (Fig.~\ref{fig:INGRID_module} right) to reject charged particles coming from outside the modules. 
The dimensions of the iron target plates are $124\times124$~$\mathrm{cm^{2}}$ in the horizontal and vertical directions and 6.5~cm along the beam direction. 
The total iron mass serving as a neutrino interaction target is 7.1~tonnes per module. 
Neutrino interaction events are selected by reconstructing the track of charged particles generated by interactions in the iron target. 
The horizontal and vertical profiles are reconstructed from the number of observed events in each module. 
The beam center is measured as the center of profile.
Each of the \kendall{eleven} tracking planes consists of 24 scintillator bars in the horizontal direction glued to 24 perpendicular bars in the vertical direction with $\mathrm{CEMEDINE^{\tiny{\textregistered}}}$ PM200. 
Each of veto planes consists of 22 scintillator bars segmented along the beam direction. 
There are 9592 scintillator bars in total. 
No iron plate was placed between the 10th and 11th tracking planes. 
\par
All the INGRID scintillator bars were produced at Fermilab~\cite{bib:Scinti}. 
The scintillator bars are made of polystyrene, infused with PPO (1\%) and POPOP (0.03\%), and are produced by extrusion in the shape of a rectangular cross section (1.0~cm $\times$ 5.0~cm). 
A white reflective coating composed of $\mathrm{TiO_2}$ infused in polystyrene is co-extruded. 
One side of the rectangular face, far from the photo-detector, is painted with $\mathrm{ELJEN^{\tiny{\textregistered}}}$ EJ-510. 
The length of the scintillator bars is 120.3~cm, 111.9~cm and 129.9~cm for tracking, top/bottom veto and right/left veto planes, respectively. 
A hole whose diameter is about 3~mm at the center of the scintillator bar allows the insertion of a wavelength shifting (WLS) fiber for light collection. 
\par
The WLS fiber, Y11(200)M by Kuraray~\cite{bib:fiber} is used for the light collection. 
The diameter of the fibers is 1.0~mm and fits the active region of the photo-detector (1.3$\times$1.3~$\mathrm{mm^{2}}$). 
The fibers are cut to the lengths of the scintillators and the cut surfaces are polished with diamond blades (\kendall{Fiberfin} Inc. FiberFin 4). 
One side of the fiber is attached to a Multi Pixel Photon Counter (MPPC, Hamamatsu S10362-13-050C)~\cite{bib:MPPC}~\cite{bib:MPPCmass} with a specially developed connector~\cite{bib:GOMIconnector}. 
The other side is painted with $\mathrm{ELJEN^{\tiny{\textregistered}}}$ EJ-510 to increase the light yield at the far side. 
\par
The MPPC signal is transported to the front-end electronics by a co-axial cable (Hirose U.FL-2LP-068).
In the electronics, the integrated charge and hit timing are digitized and recorded with a 2.5~photo-electron (PE) threshold. 
The bias voltage applied to the MPPCs is controlled by the front-end electronics with a precision of 0.02~V. 
A detailed description of the electronics can be found in~\cite{bib:TFB}. 
\par
\begin{figure}[htbp]
 \begin{center}
  \includegraphics[width=0.7\textwidth]{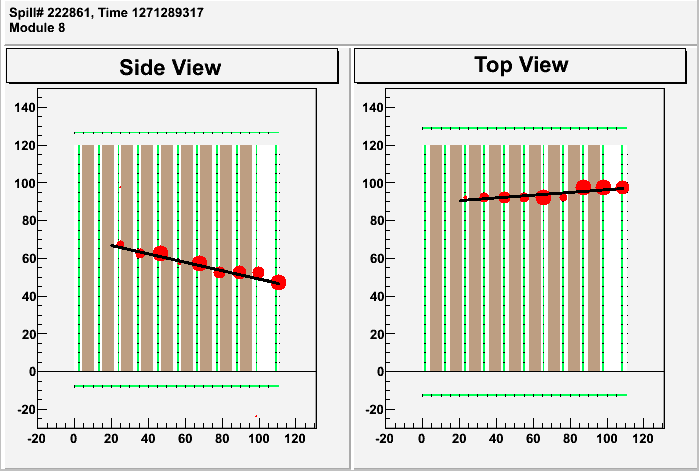}
  \caption{\modif{Typical neutrino interaction event candidate in one of the modules. A beam neutrino enters from the left. The size of the circles is proportional to the observed number of PE at scintillator bars, and black lines show the reconstructed tracks.} }
  \label{fig:typical_event}
 \end{center}
\end{figure}

\modif{Figure \ref{fig:typical_event} shows an example of neutrino event candidates in one of the modules. }
\par
\section{Basic performance of the detector}
INGRID identifies neutrino events by detecting tracks from muons. 
The hit efficiency for muon tracks is monitored with the beam induced muons incoming from outside of the detector. 
The mean light yield and timing resolution are monitored with cosmic-ray data. 
\subsection{Mean light yield}\label{subsec:ly}
\begin{figure}[htbp]
 \begin{center}
  \includegraphics[width=0.7\textwidth]{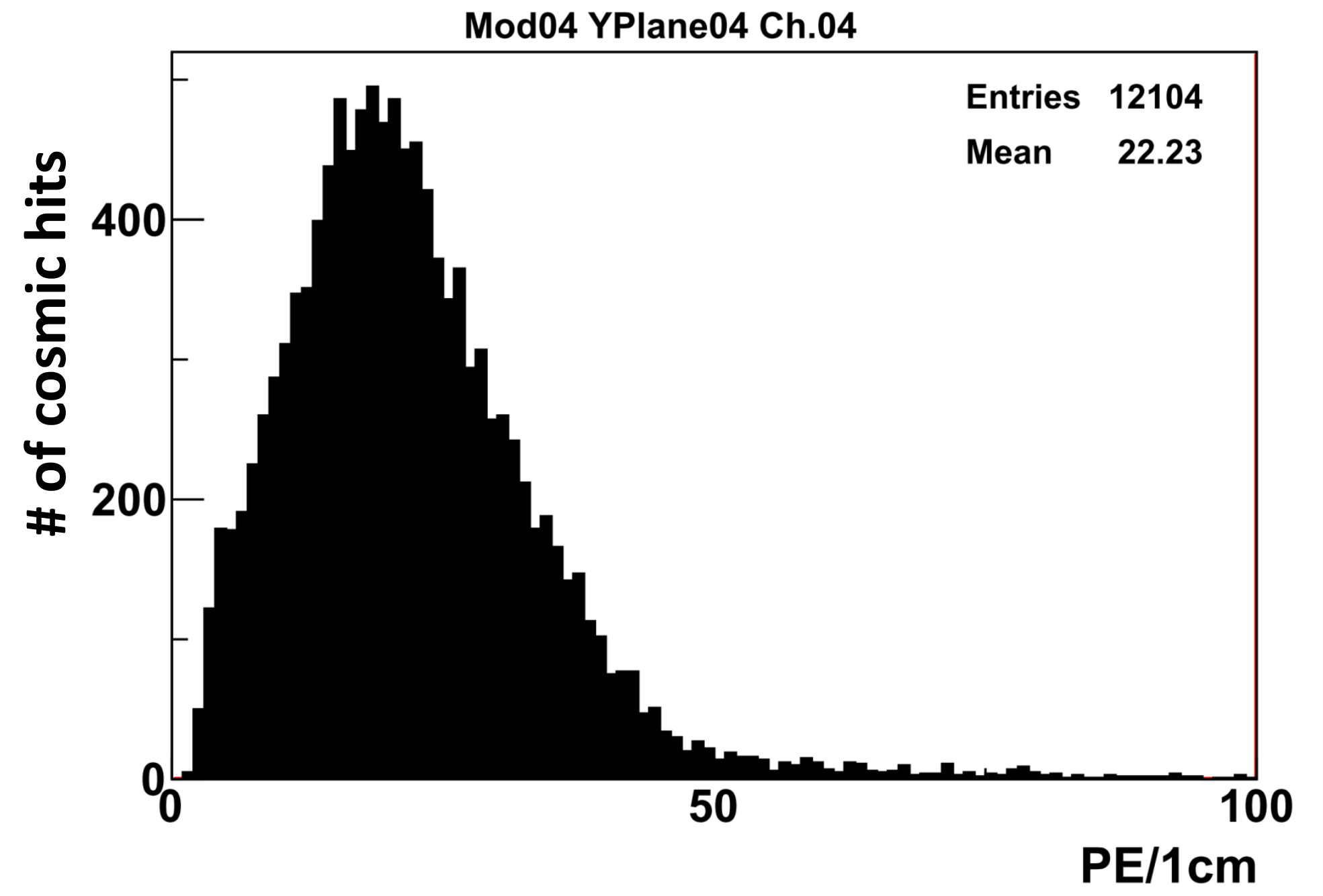}
  \caption{Typical light yield distribution. Light yield is normalized by the path length. }
  \label{fig:LY}
 \end{center}
\end{figure}

\begin{figure}[htbp]
 \begin{center}
  \includegraphics[width=0.7\textwidth]{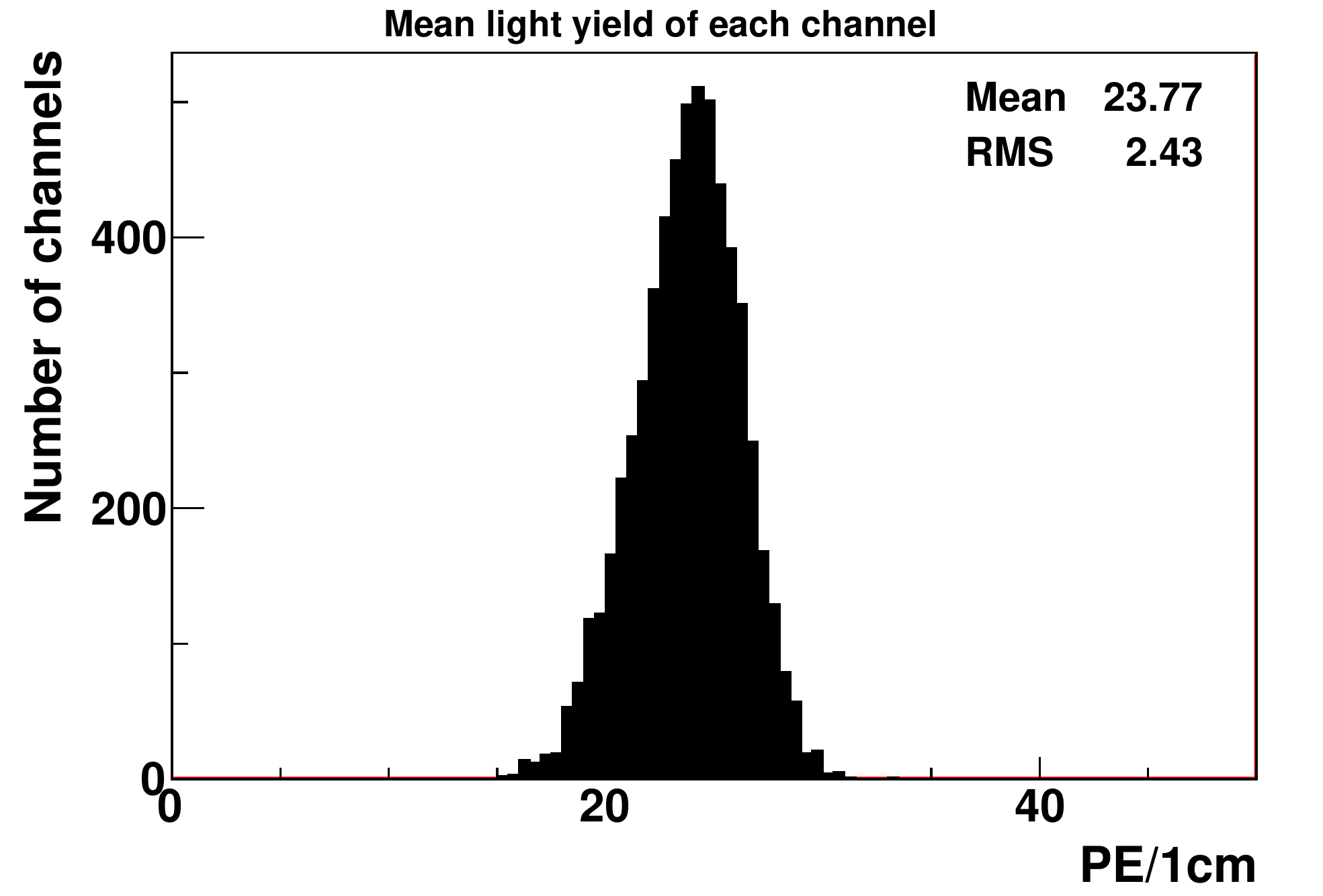}
  \caption{Mean light yields. }
  \label{fig:LYall}
 \end{center}
\end{figure}

The mean light yield per 1cm of a muon track is monitored with inter-spill cosmic-ray data for each \kendall{scintillator bar}. 
A typical light yield distribution of one channel is shown in Fig.~\ref{fig:LY}. 
The distribution is consistent with the Landau distribution. 
The mean light \kendall{yields} of all the channels are shown in Fig.~\ref{fig:LYall}. 
\modif{The average mean light yield normalized to unit length is 24 PE/cm. }
\subsection{Hit efficiency}
The hit efficiency is evaluated by muons with the following procedure; 
the muon track is reconstructed without using the hit information in the scintillator plane being evaluated, and then in that plane, channels expected to have hits from the track trajectory are checked whether they have \kendall{a} hit or not. 
\par
\begin{figure}[htbp]
 \begin{center}
  \includegraphics[width=0.7\textwidth]{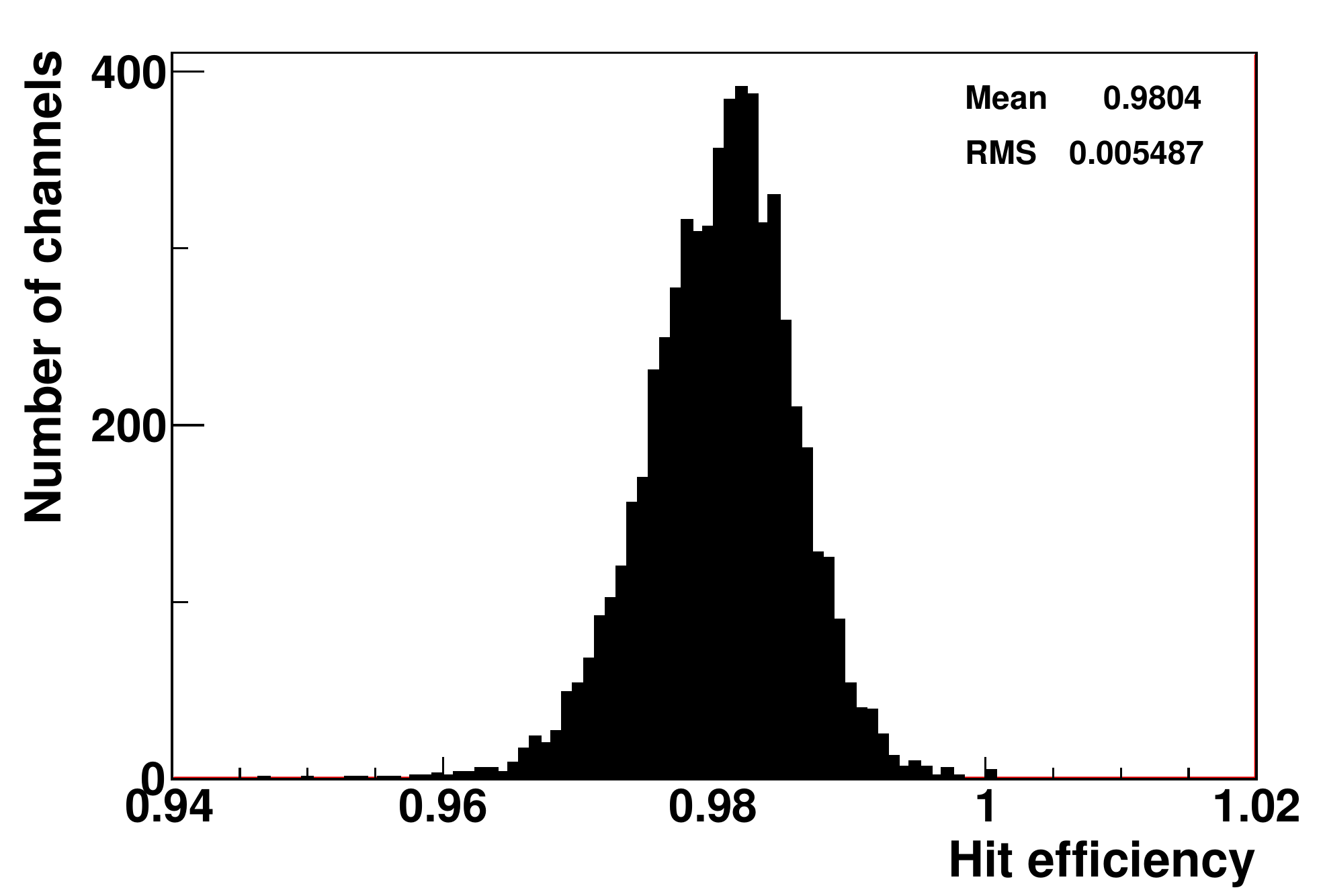}
  \caption{Hit efficiency for all channels.}
  \label{fig:cheff}
 \end{center}
\end{figure}
\begin{figure}[htbp]
 \begin{center}
  \includegraphics[width=0.7\textwidth]{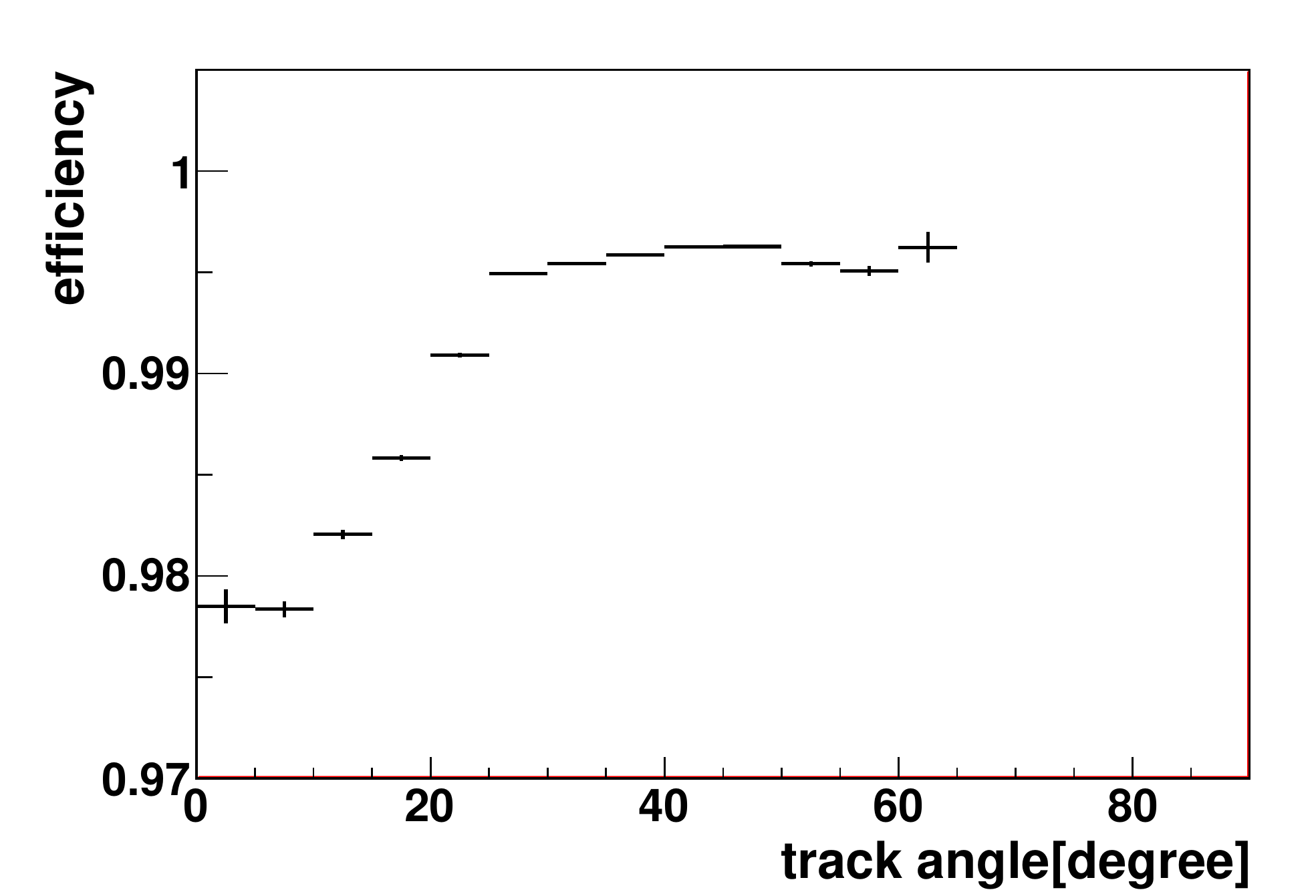}
  \caption{Hit efficiency as a function of reconstructed track angle measured by cosmic-ray data. }
  \label{fig:eff_angle}
 \end{center}
\end{figure}

Figure \ref{fig:cheff} shows the result of the hit efficiency measurement with the beam induced muons. 
Figure~\ref{fig:eff_angle} shows the efficiency as a function of the track angle measured using cosmic-ray muons. 
The track angle is defined as the angle between the designed beam direction and the reconstructed track. 
\kendall{The main reason for the inefficiency is because of the small gap between scintillator bars, so the efficiency depends on track angle; a particle with small angle has more probability to go through the gap. }
As a result, the efficiency is smaller than that expected from PE statistics with measured light yield described in section \ref{subsec:ly}. 
\subsection{Hit timing resolution}
\begin{figure}[htbp]
 \begin{center}
  \includegraphics[width=0.65\textwidth]{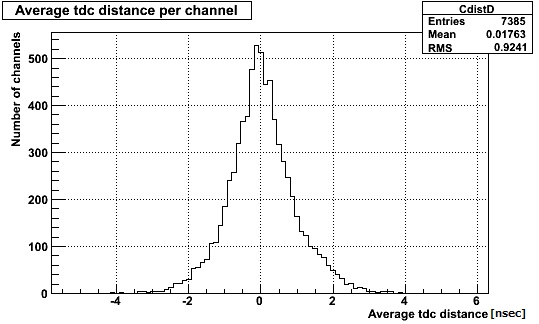}
  \caption{Time difference of hit channels from the average hit times for cosmic-ray tracks.}
  \label{fig:time_diff}
 \end{center}
\end{figure}

Hit timing resolution is estimated by measuring the time difference among hit channels for cosmic-ray tracks. 
Figure~\ref{fig:time_diff} shows the time differences of each hit channel from the average of all channels after the correction for differences in the readout cable length and the light propagation time through the fiber. 
The RMS is 0.9~nsec, which corresponds to the timing resolution if all the channels have the same resolution. 
The width of the primary proton beam bunch was about 30 nsec during RUN~1 and 2, so this resolution is sufficient for selecting the beam events. 
\section{Monte Carlo simulations for the neutrino beam measurements.}\label{sec:mc}
The Monte Carlo (MC) simulations for the neutrino beam measurements \kendall{consist} of three main parts. 
The first is a simulation of the neutrino beam production, which predicts the neutrino flux and energy spectrum of each neutrino flavor. 
The second is a neutrino interaction simulation, which calculates the cross-section of the neutrino interaction and kinematics of final state particles taking into account the intranuclear interactions of hadrons. 
The third step is a detector response simulation to reproduce the scintillator light yield, and the response of WLS fibers and MPPC. 
\subsection{Neutrino beam prediction}
\begin{figure}[htbp]
  \begin{center}
   \includegraphics[width=0.65\textwidth]{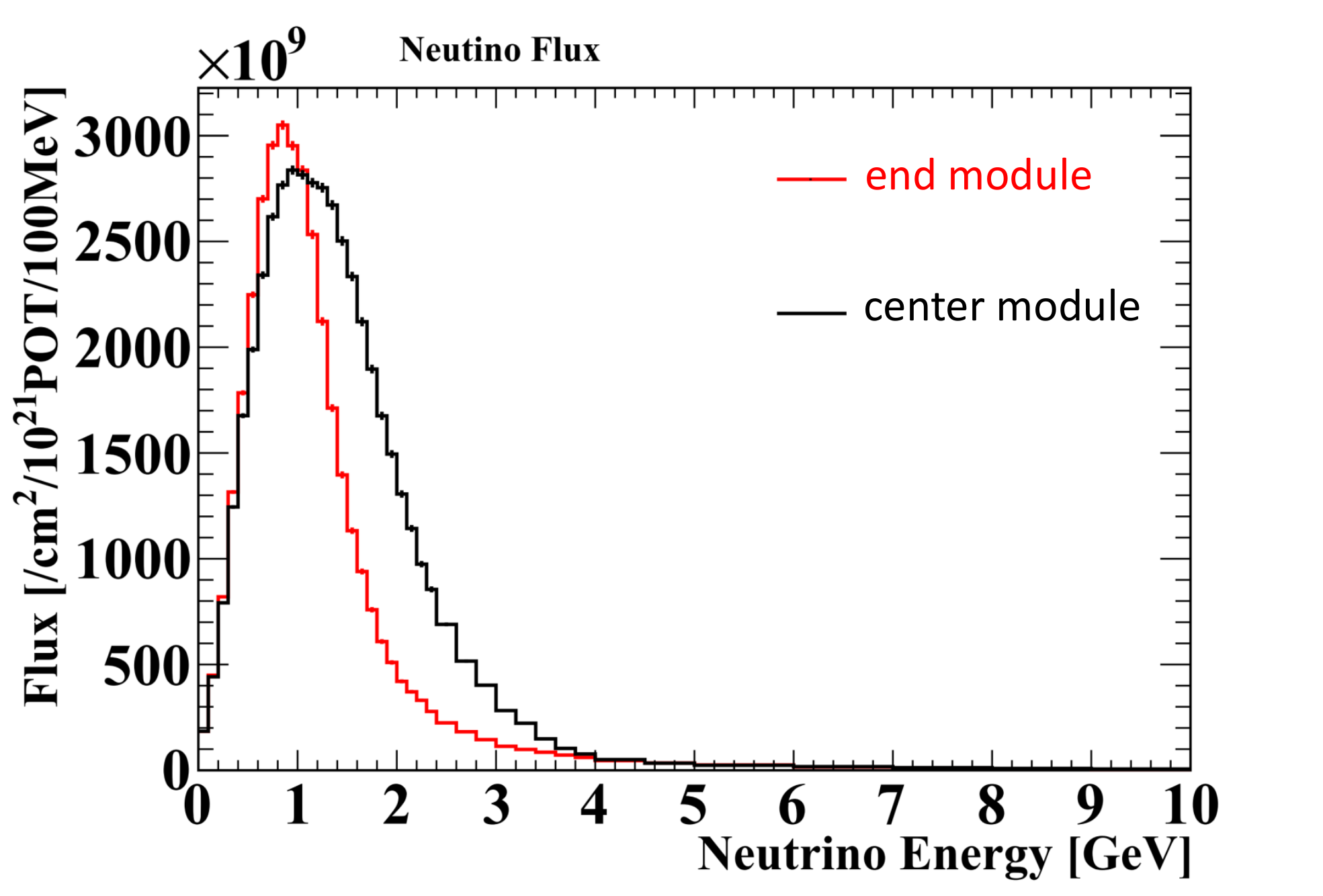}
   \caption{Neutrino energy spectrum predicted by JNUBEAM at the center and end modules.}
   \label{fig:nu_spectrum}
  \end{center}
\end{figure}

To predict neutrino fluxes and energy spectra, a neutrino beam Monte Carlo simulation, called JNUBEAM~\cite{bib:T2KNIM}, was developed based on the GEANT3 framework~\cite{bib:GEANT}. 
\modif{We compute the neutrino beam fluxes starting from models (FLUKA~\cite{bib:FLUKA, bib:FLUKA2} and GCALOR~\cite{bib:GCALOR}) and tuning them to experimental data (NA61/SHINE~\cite{bib:NA61} and Eichten \textit{et al.}~\cite{bib:Eichten}). }
Energy spectra at the center and end of the horizontal modules are shown in Fig.~\ref{fig:nu_spectrum}. 
Because each module covers a different off-axis angle, the neutrino energy spectrum at each module location is different. 
The difference \kendall{in} the average neutrino energy between the center module and the end module is about 0.2~GeV. 
Energy spectra at 10~m upstream from INGRID are predicted with the same procedure in order to simulate the background events from neutrino interactions in the wall of the experimental hall.  
\subsection{Neutrino interaction simulation}
Neutrino interactions with iron in INGRID are simulated using the NEUT program libraries~\cite{bib:NEUT}. 
The neutrino interactions in the scintillator tracker are generated, taking into account the mass ratio. 
Combined with the neutrino flux predicted by JNUBEAM, \modif{we expect about $4\times10^{7}$ neutrino interactions for $10^{21}$ POT in the 14 modules consisting of horizontal and vertical rows. }
For the background event simulation, the neutrino interactions on an equal mix of carbon and hydrogen are generated uniformly in the upstream wall of the experimental hall. 
The wall mainly consists of \kendall{sand}, but for simplicity carbon and hydrogen are used in the simulation. 
The number of generated interactions is normalized by the beam induced muon events (described in Sec.\ref{sec:ana_veto}) in data. 
\subsection{INGRID detector response simulation}
Detector response simulation was developed with the Geant4 framework~\cite{bib:GEANT}. 
The simulation includes a detailed geometry of the experimental hall. 
\par
The energy deposit in each scintillator bar is simulated by Geant4 library and is converted to the number of PE at each MPPC. 
The conversion factor from the energy deposit to a number of PE is determined based on the measured light yields with cosmic-rays. 
The cross section of the scintillator bar is tuned to reproduce the hit inefficiency due to the dead region of the bar. 
\kendall{The} scintillator quenching effect is simulated using Birk's law with the value measured in~\cite{bib:scibar_attenuation}. 
Attenuation in the fiber is taken into account based on the measured attenuation length~\cite{bib:test_beam}.
The response of MPPC, such as saturation due to the finite number of photo-diodes, is modeled based on test bench measurements~\cite{bib:MPPC_T2K}. 
\par
The dimensions and mass of the iron target plates are implemented with the design value. 
For analysis, we make a correction for the measured mass difference for each iron plate, as described in Sec.\ref{subsec:correction}. 
\par
\section{Neutrino event selection}\label{sec:analysis}
The neutrino beam profile is reconstructed from the number of neutrino interaction events at each module. 
This section describes the selection procedure for neutrino interactions and the systematic error.  
\par
\subsection{Selection criteria}\label{sec:selection}
A neutrino interaction event is identified by a long track from a charged particle generated by the neutrino interaction. 
First, pre-selections are applied to reject accidental noise events. 
Then, tracks are reconstructed using hit information. 
After that, charged particles from outside of the module are rejected with the veto planes and the reconstructed event vertex is required to be inside the fiducial volume (FV). 
In these selections, each module is treated separately. 
The event selection criteria are described in the following sections. 
\subsubsection{Event definition}
When there are four or more hits in a 100~nsec time window, all hits within $\pm50$~nsec compose an event. 
\subsubsection{Pre-selections}
\begin{figure}[htbp]
 \begin{center}
  \includegraphics[width=0.7\textwidth]{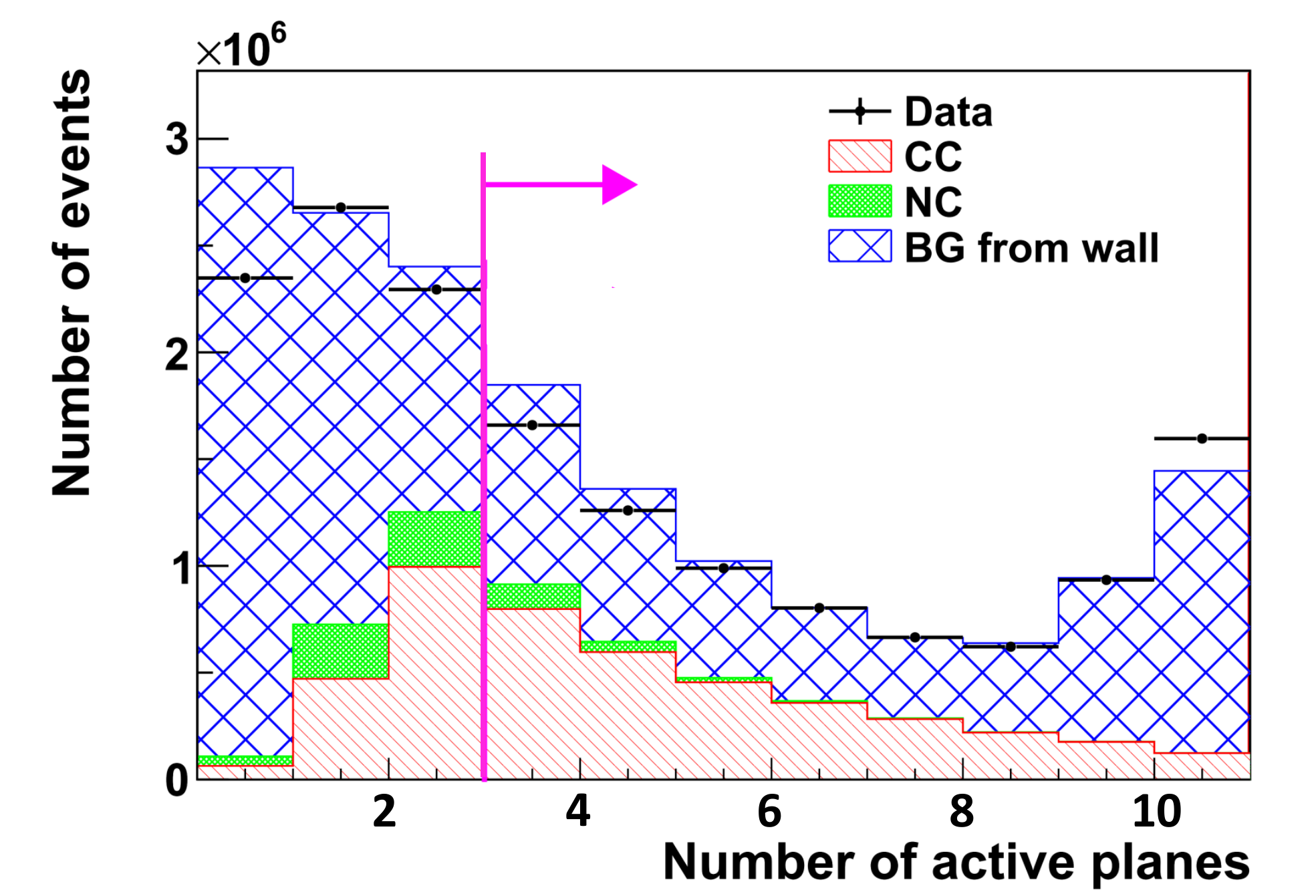}
  \caption{Number of active planes. Events with more than two active planes are selected. }
  \label{fig:selection_actpln}
 \end{center}
\end{figure}

\begin{figure}[htbp]
\begin{minipage}{0.475\hsize}
 \begin{center}
  \includegraphics[width=0.9\textwidth]{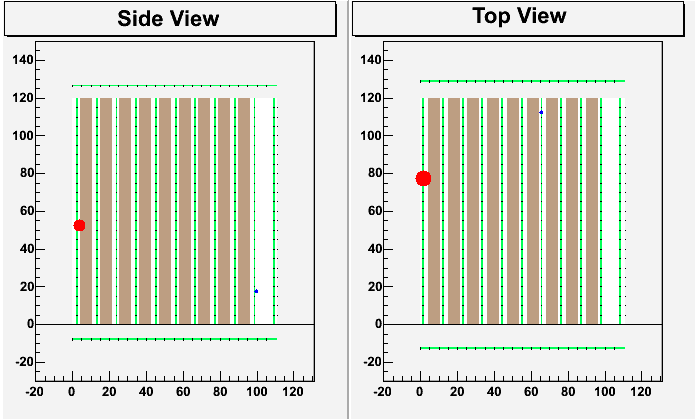}
 \end{center}
 \end{minipage}
 \begin{minipage}{0.05\hsize}
 \end{minipage}
 \begin{minipage}{0.475\hsize}
 \begin{center}
  \includegraphics[width=0.9\textwidth]{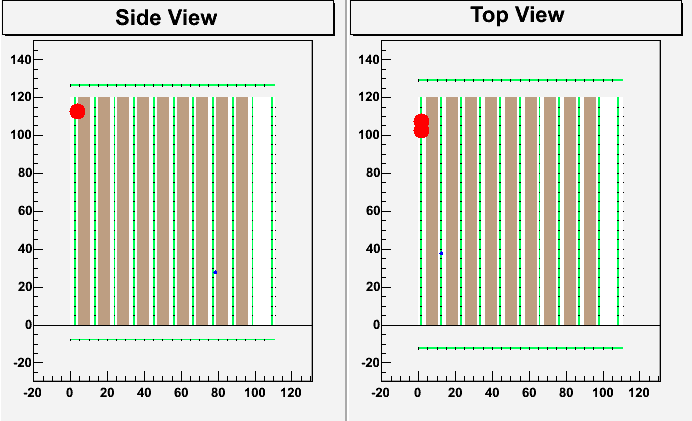}
 \end{center}
 \end{minipage}
 \caption{\modif{Examples of the MC event with no active plane. The red circle shows a hit by the particle and the blue circle shows the hit by MPPC noise. }}
 \label{fig:nact1}
\end{figure}

A tracking plane with at least one hit in both x and y layers is defined as an "active" plane. 
We use a right hand coordinate in which z is along the beam direction and y is upward in the vertical direction. 
Events with three or more active planes are selected as shown in Fig.~\ref{fig:selection_actpln}. 
\modif{There is a discrepancy in the number of events with no active plane between data and MC. 
Figure \ref{fig:nact1} shows a MC event display of a typical event with no active plane. 
The first tracking plane is not counted as the active planes, but are used fro veto. 
Most of the events originate in low energy particles produced by the neutrino interaction in the wall of the experimental hall. 
We consider that the discrepancy is caused by uncertainties to simulate these low energy particles. 
}
\par
\begin{figure}
 \begin{center}
  \includegraphics[width=0.7\textwidth]{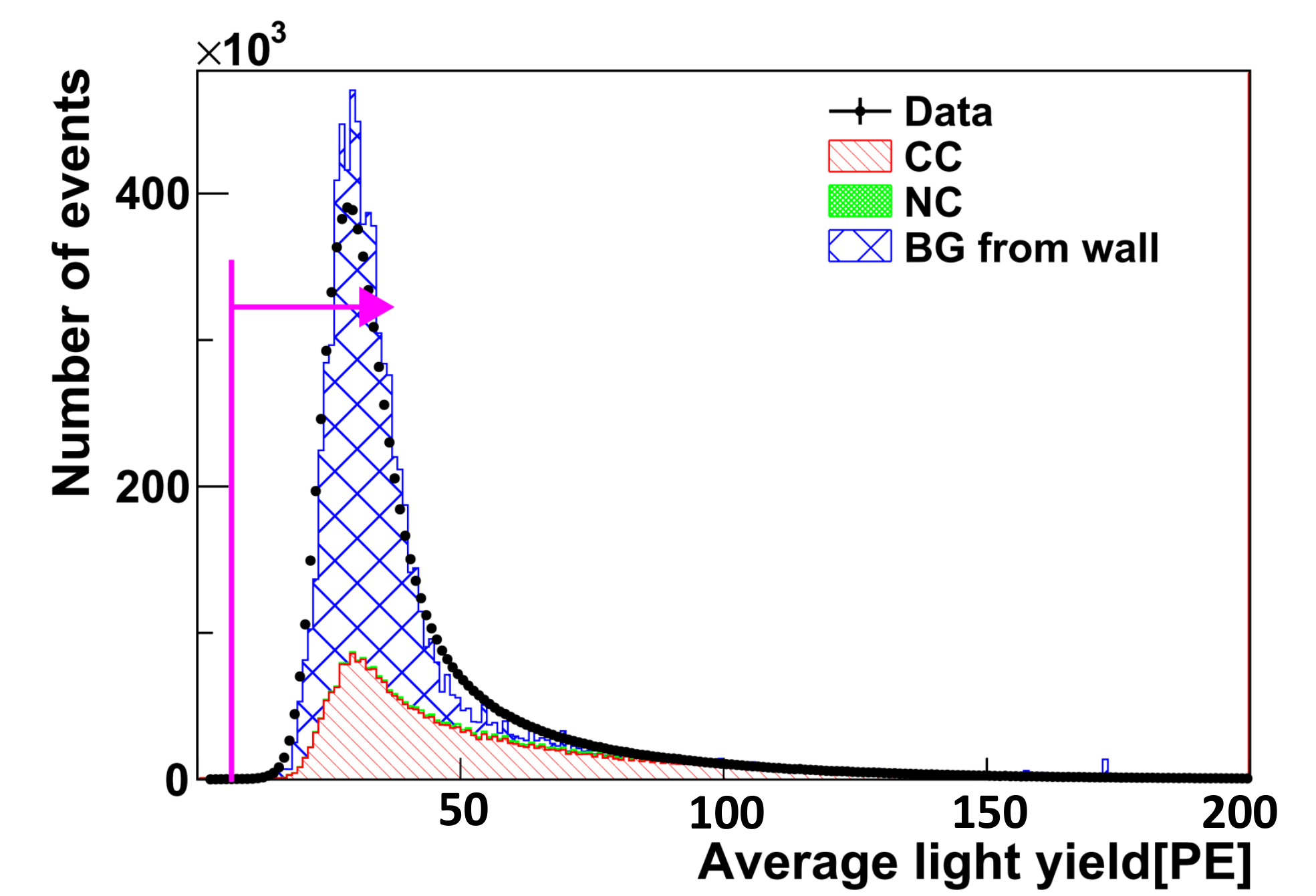}
  \caption{Light yield averaged over active layers. Events with more than 6.5~PE are selected. }
  \label{fig:selection_layerpe}
 \end{center}
\end{figure}

After the selection with the number of active planes, light yields averaged over x or y layers of the active planes are required to be larger than 6.5 PE for both x layers and y layers as shown in Fig.\ref{fig:selection_layerpe}; inefficiency due to this selection is negligible for muon tracks. 
\subsubsection{Tracking}
\begin{figure}[htbp]
\begin{minipage}{0.475\hsize}
 \begin{center}
  \includegraphics[width=0.9\textwidth]{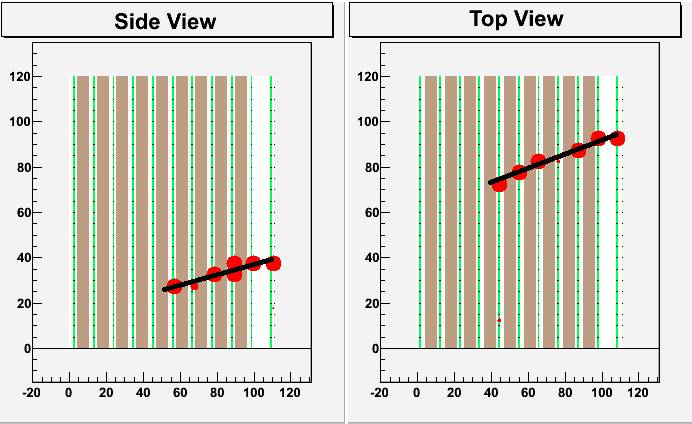}
 \end{center}
 \end{minipage}
 \begin{minipage}{0.05\hsize}
 \end{minipage}
 \begin{minipage}{0.475\hsize}
 \begin{center}
  \includegraphics[width=0.9\textwidth]{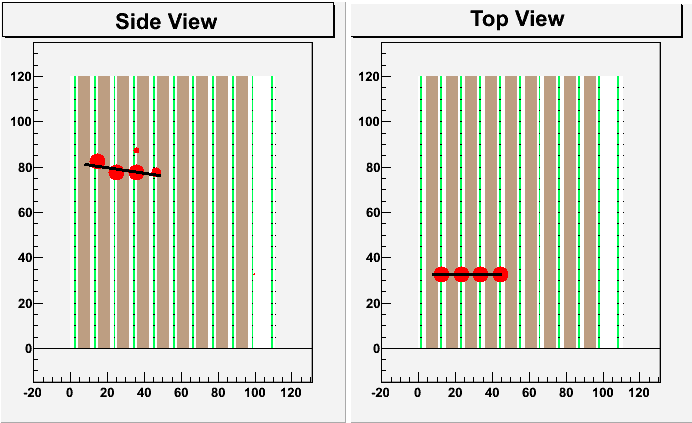}
 \end{center}
 \end{minipage}
 \caption{Examples of the reconstructed tracks. The size of the circles shows the observed number of PE at scintillator bars, and black lines show the reconstructed tracks. }
 \label{fig:selection_TRKexample}
\end{figure}

After the previous selections, tracks in x-z and y-z projection are reconstructed independently with a simple algorithm;
first, hits in the most downstream active layer are adopted \kendall{as} the end point of the track. 
Then the track is extrapolated to upstream layers by checking the upstream hits. 
A hit is included in the track if the hit position is within two scintillator bars from the straight line extrapolated from the downstream hits. 
Figure~\ref{fig:selection_TRKexample} shows an example of a reconstructed track. 
Tracking efficiency is checked with cosmic-ray data and the efficiency is $\sim95$\% for cosmic-rays passing three scintillator planes. 
\par
\begin{figure}[htbp]
 \begin{center}
  \includegraphics[width=0.7\textwidth]{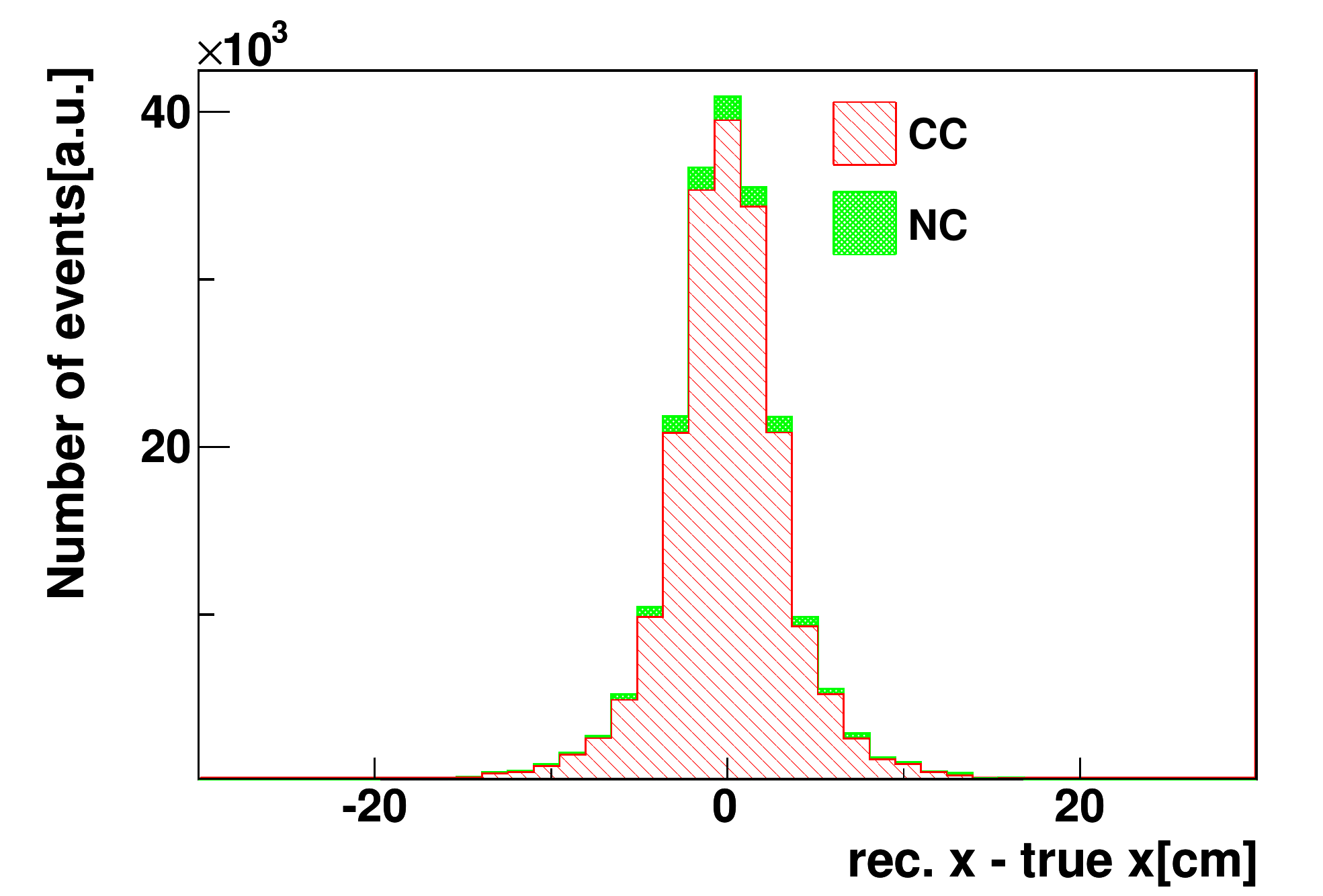}
  \caption{Differences between true and reconstructed vertex position in the x direction for MC events.}
  \label{fig:resolution_verx}
 \end{center}
\end{figure}
\begin{figure}[htbp]
 \begin{center}
  \includegraphics[width=0.7\textwidth]{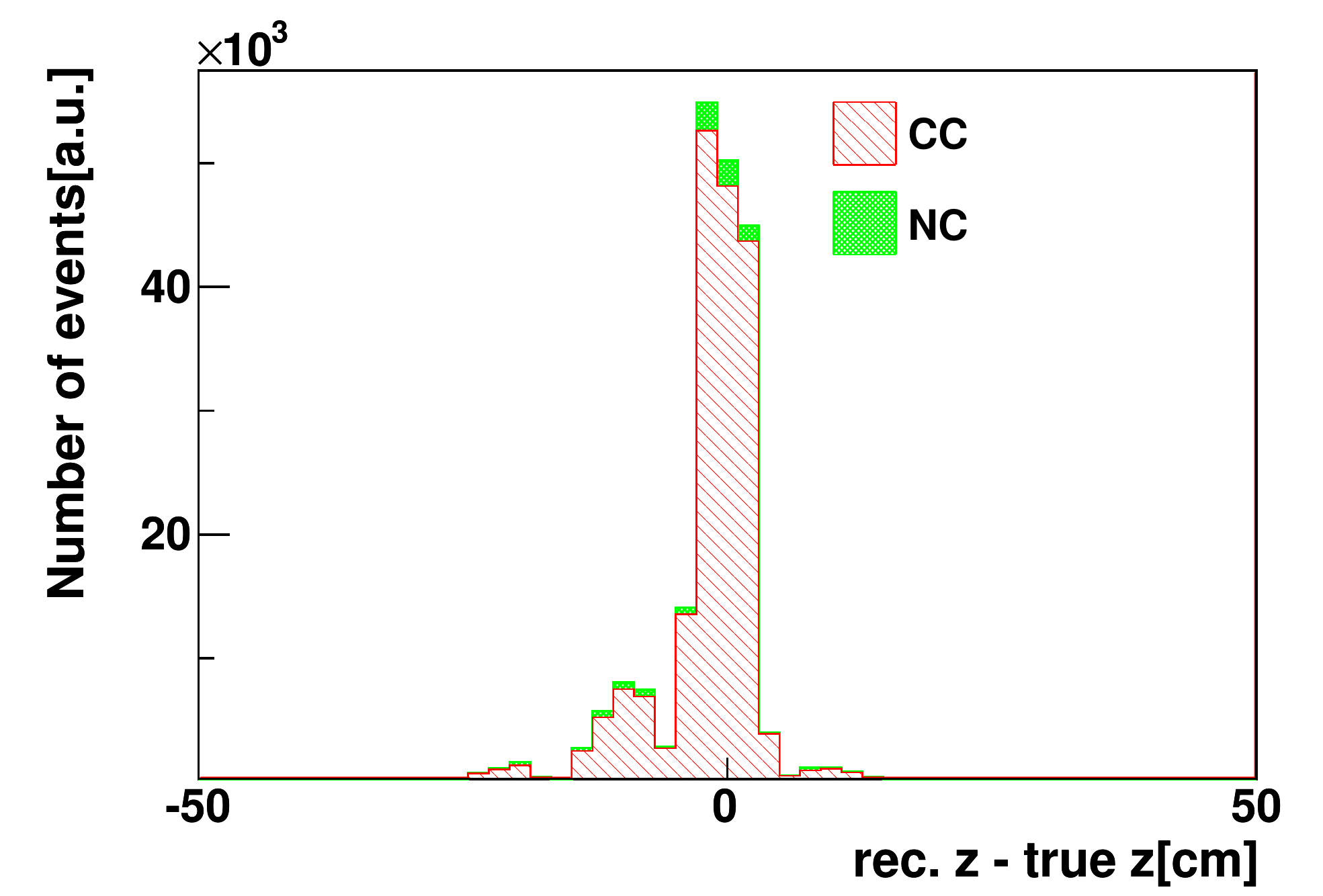}
  \caption{Differences between true and reconstructed vertex position in the z direction for MC events.}
  \label{fig:resolution_verz}
 \end{center}
\end{figure}
\begin{figure}
 \begin{center}
  \includegraphics[width=0.7\textwidth]{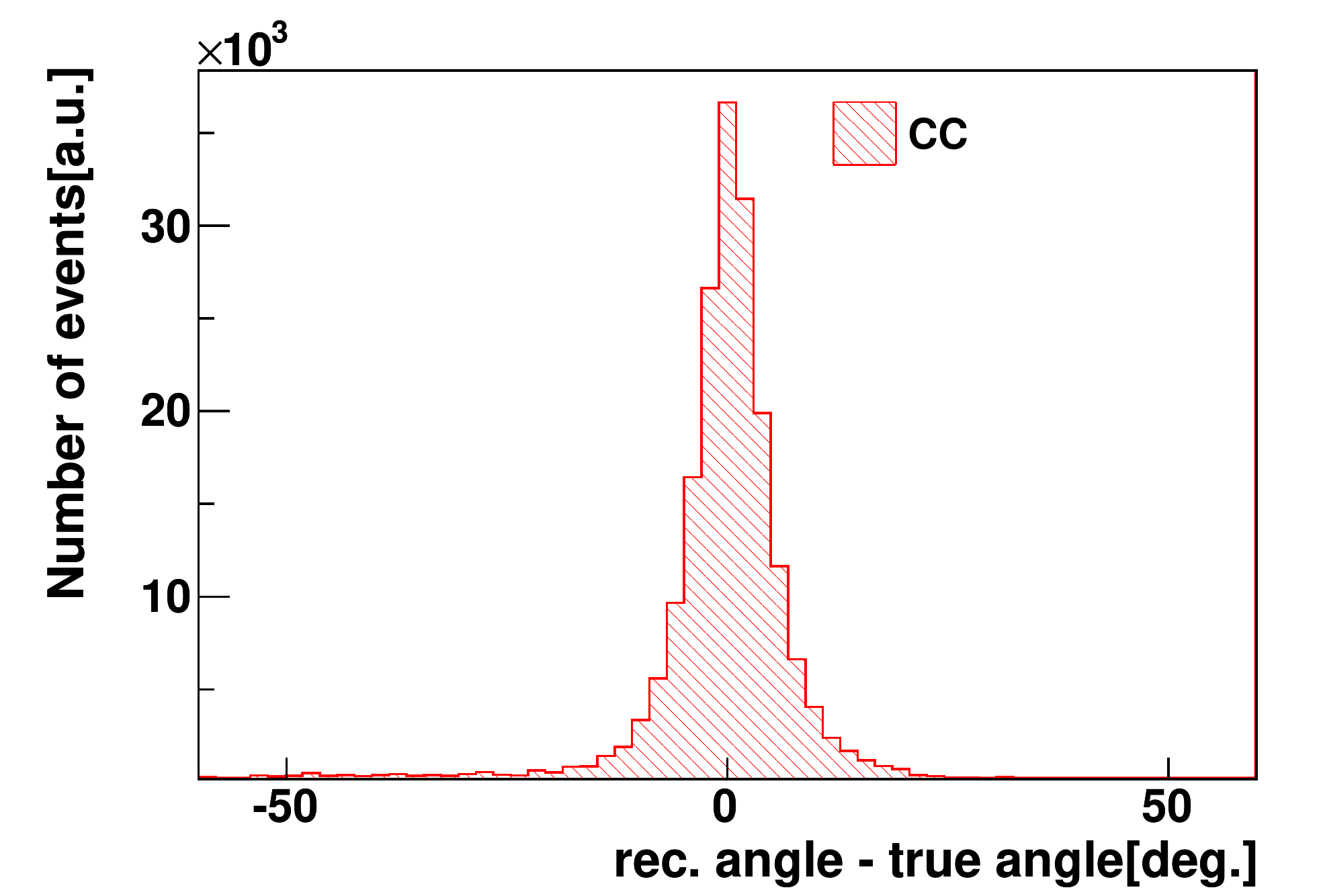}
  \caption{Angles between true and reconstructed tracks.}
  \label{fig:resolution_angle}
 \end{center}
\end{figure}

The vertex position is reconstructed as the most upstream point of the track for each projection. 
The track angle is obtained by fitting the hits composing the track with a straight line. 
Figures~\ref{fig:resolution_verx} and \ref{fig:resolution_verz} show differences between true and reconstructed x and z vertices, respectively, for MC events.  
The RMS for the x vertices is 2.7~cm. 
Figure~\ref{fig:resolution_angle} shows the distribution of 3D angle between true and reconstructed muon tracks for MC events. 
The RMS is 3.8~degrees. 
\par
\begin{figure}[htbp]
 \begin{center}
  \includegraphics[width=0.7\textwidth]{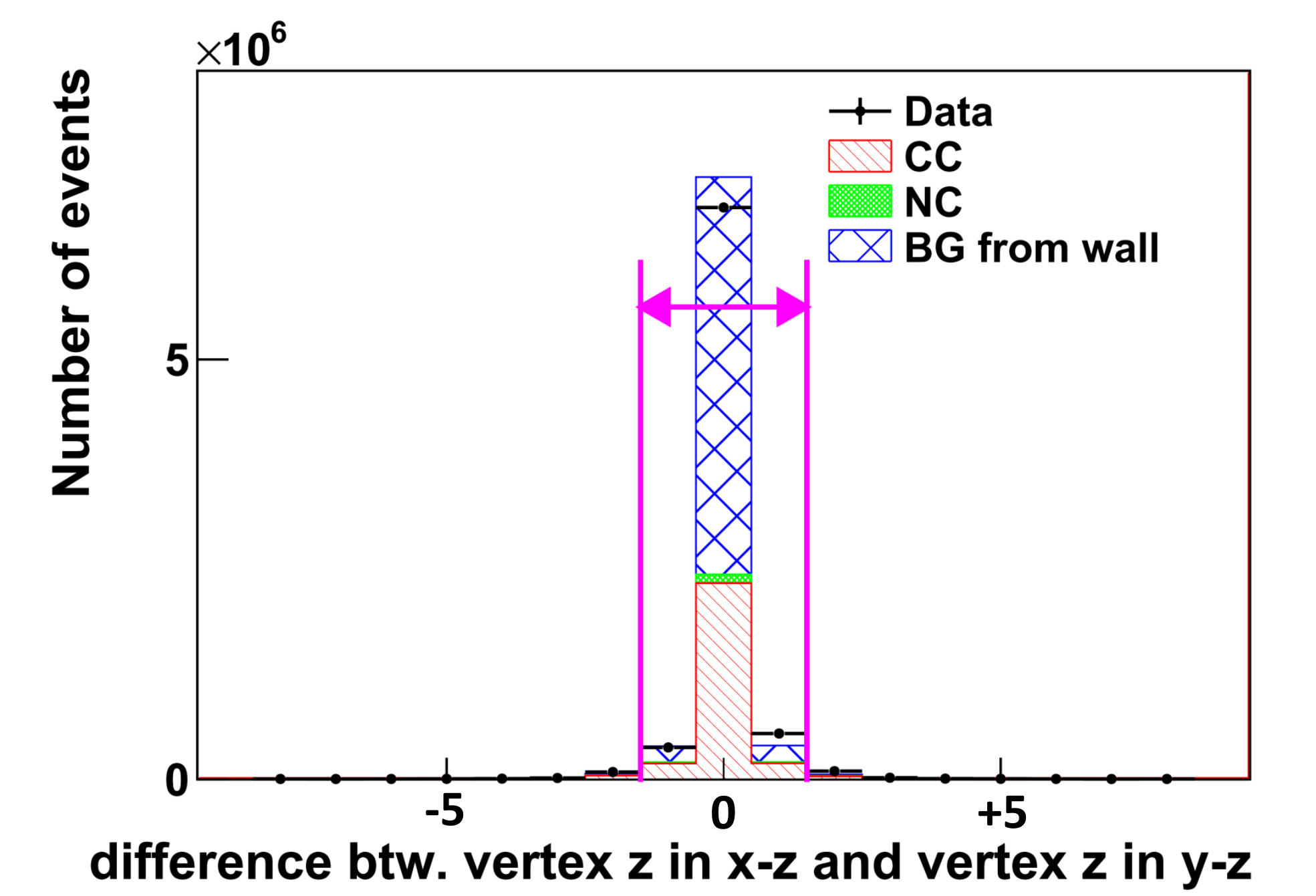}
  \caption{Difference of the z vertex position between x-z and y-z projections.}
  \label{fig:selection_trkmatch}
 \end{center}
\end{figure}

After the tracking, some badly fitted tracks are rejected by using the position difference of the vertex z between x-z and y-z projections. 
The difference is required to be within $\pm1$ plane, as shown in Fig.~\ref{fig:selection_trkmatch}. 
\subsubsection{Timing cut}
\begin{figure}
 \begin{center}
  \includegraphics[width=0.7\textwidth]{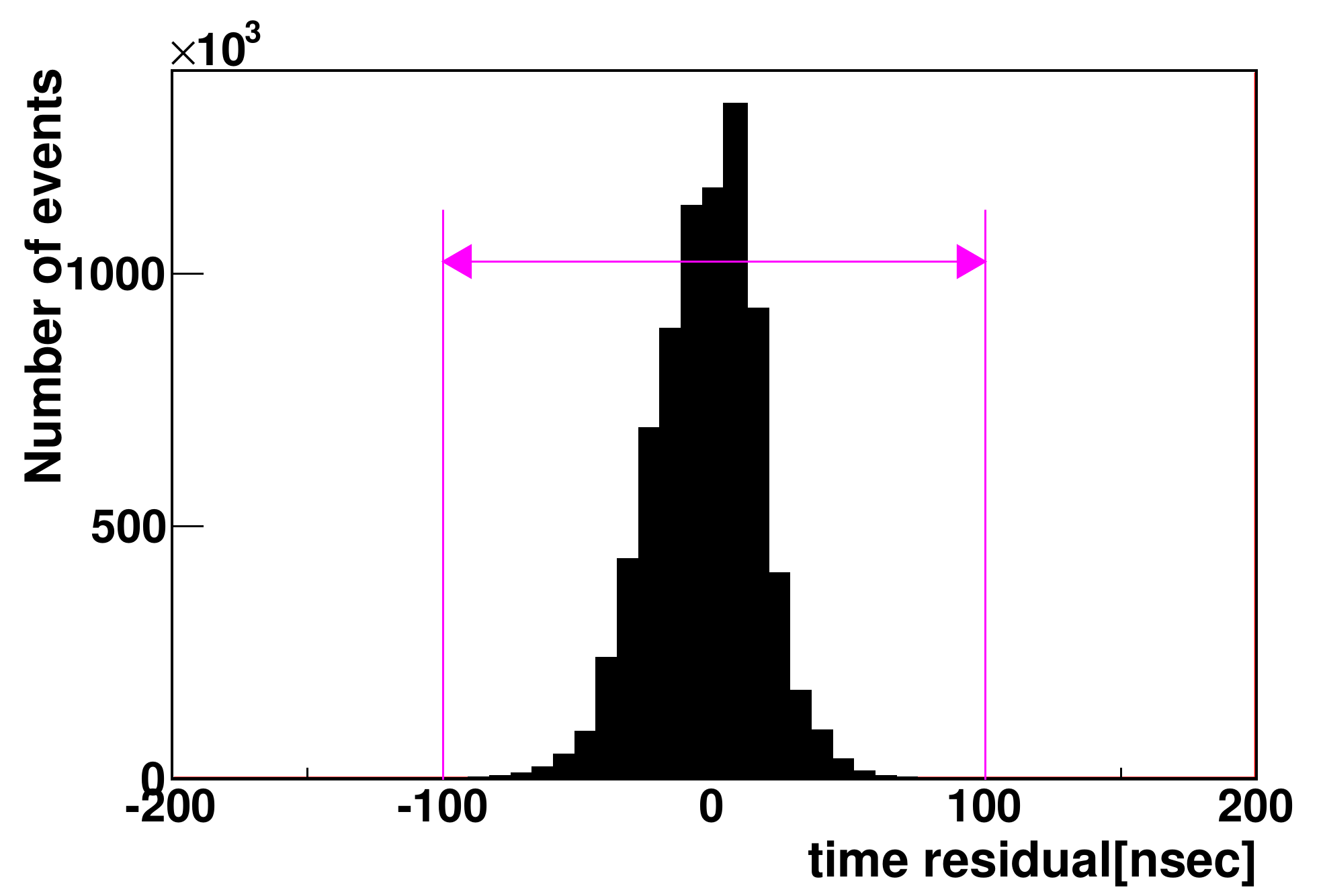}
  \caption{
Time difference between measured event timing and expected neutrino event timing. 
Events within $\pm100$ nsec are selected.
}
  \label{fig:selection_timing}
 \end{center}
\end{figure}

To reject off-timing events such as cosmic-ray events, only events of $\pm100$ nsec from the expected timing are selected (Fig.\ref{fig:selection_timing}). 
The expected timing is evaluated with the primary proton beam timing~\cite{bib:T2KNIM}, the time of flight of the particles from the target to INGRID, and the delay \kendall{of} the electronics and cables. 
The event timing is defined by the hit at the start point of the track. 
\subsubsection{\kendall{Veto} and Fiducial Volume (FV) cuts}\label{sec:ana_veto}
\begin{figure}[htbp]
\begin{minipage}{0.475\hsize}
 \begin{center}
  \includegraphics[width=0.95\textwidth]{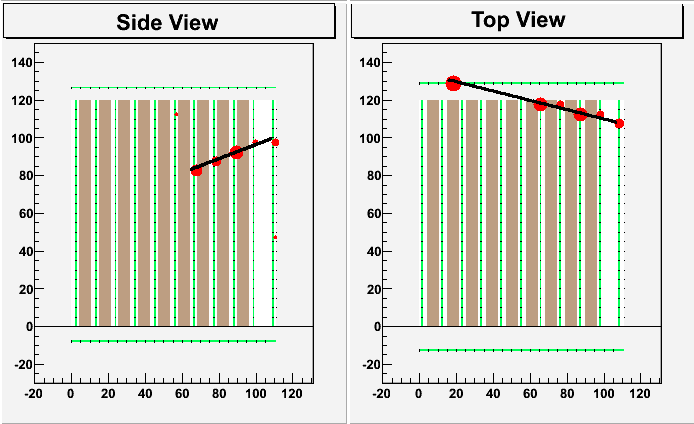}
 \end{center}
 \end{minipage}
 \begin{minipage}{0.05\hsize}
 \end{minipage}
 \begin{minipage}{0.475\hsize}
 \begin{center}
  \includegraphics[width=0.95\textwidth]{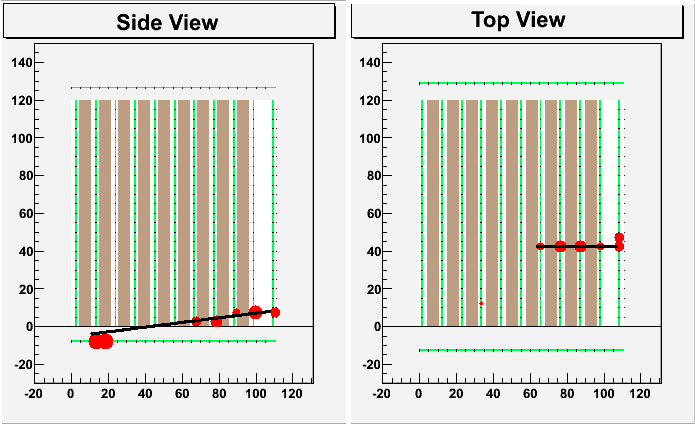}
 \end{center}
 \end{minipage}
 \caption{Event displays of rejected events by the veto cut.}
 \label{fig:BGexample}
\end{figure}

\begin{figure}[htbp]
\begin{minipage}{0.475\hsize}
 \begin{center}
  \includegraphics[width=0.9\textwidth]{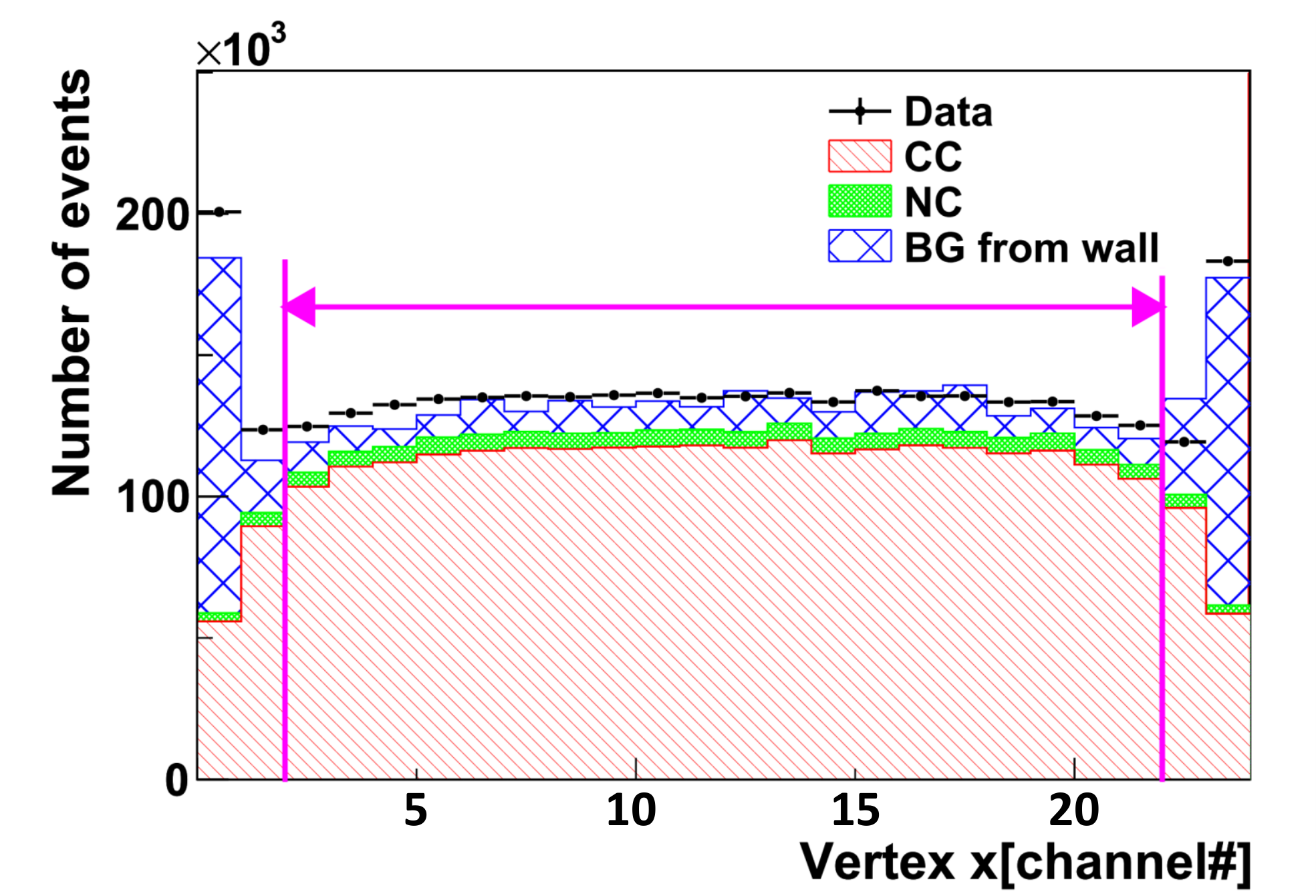}
 \end{center}
\end{minipage}
\begin{minipage}{0.05\hsize}
\end{minipage}
\begin{minipage}{0.475\hsize}
 \begin{center}
  \includegraphics[width=0.9\textwidth]{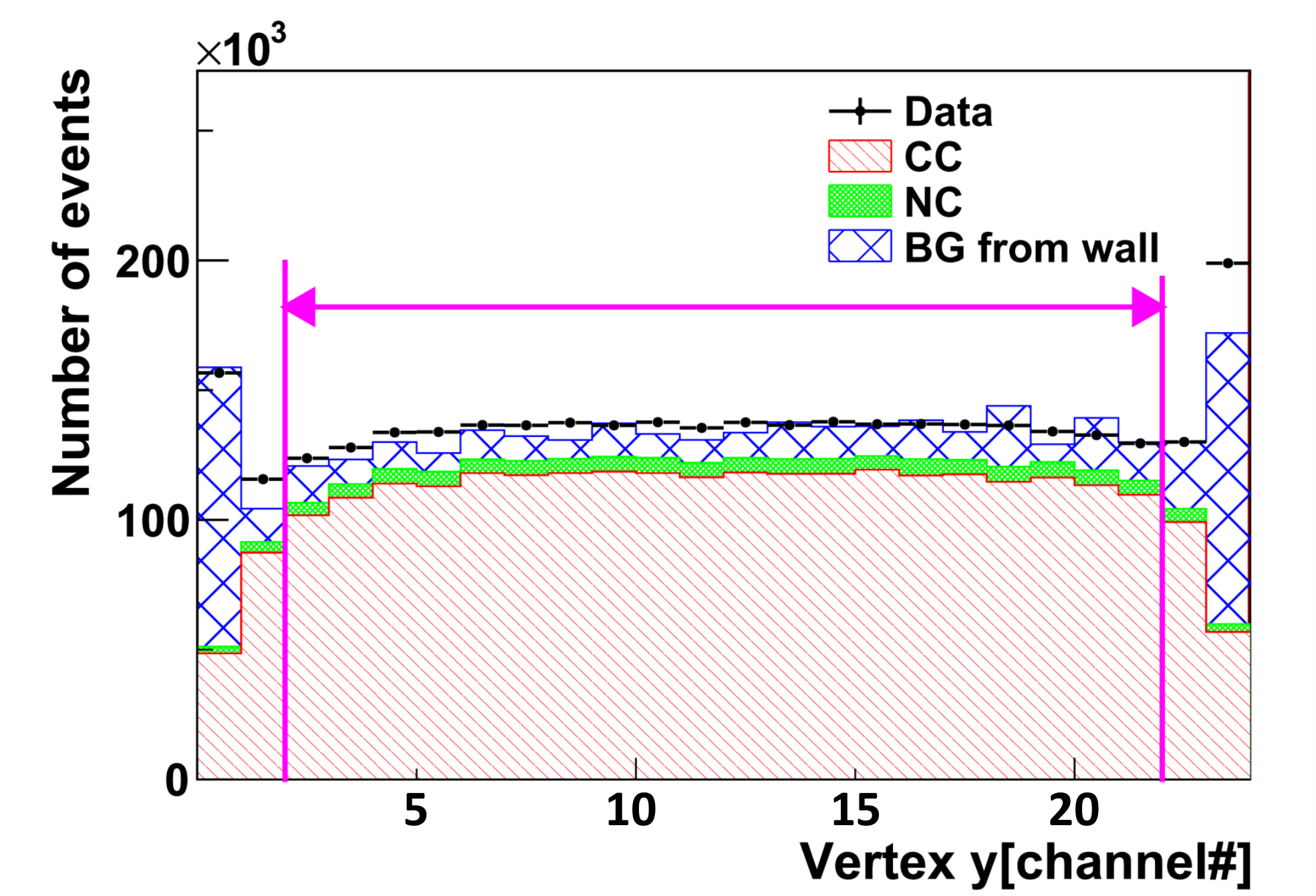}
 \end{center}
 \end{minipage}
 \caption{Vertex x and y distributions. A volume composed of the 3rd to 22nd scintillator bars in the x and y directions is defined as Fiducial Volume (FV).}
 \label{fig:verx_very}
\end{figure}

Two selections are applied to reject incoming particles produced by neutrino interactions in the upstream material, such as the wall of the experimental hall. 
First, events which have a hit in a veto plane or the first tracker plane at the \kendall{upstream position extrapolated from} the reconstructed track are rejected. 
Event displays of events rejected by the veto cut are shown in Fig.~\ref{fig:BGexample}. 
After the veto cut, the fiducial volume (FV) cut is applied. 
The FV of each module is defined as a volume composed \kendall{of} the 3rd to 22nd of the 24 scintillator bars in the x and y directions, and from \kendall{the} second to \kendall{the} ninth tracker plane in the z direction.   
Events having a vertex inside the FV are selected as shown in Fig.~\ref{fig:verx_very}. 
The events rejected by these selections are identified as `beam induced muon` events. \par
\begin{figure}
\begin{minipage}{0.475\hsize}
 \begin{center}
  \includegraphics[width=0.9\textwidth]{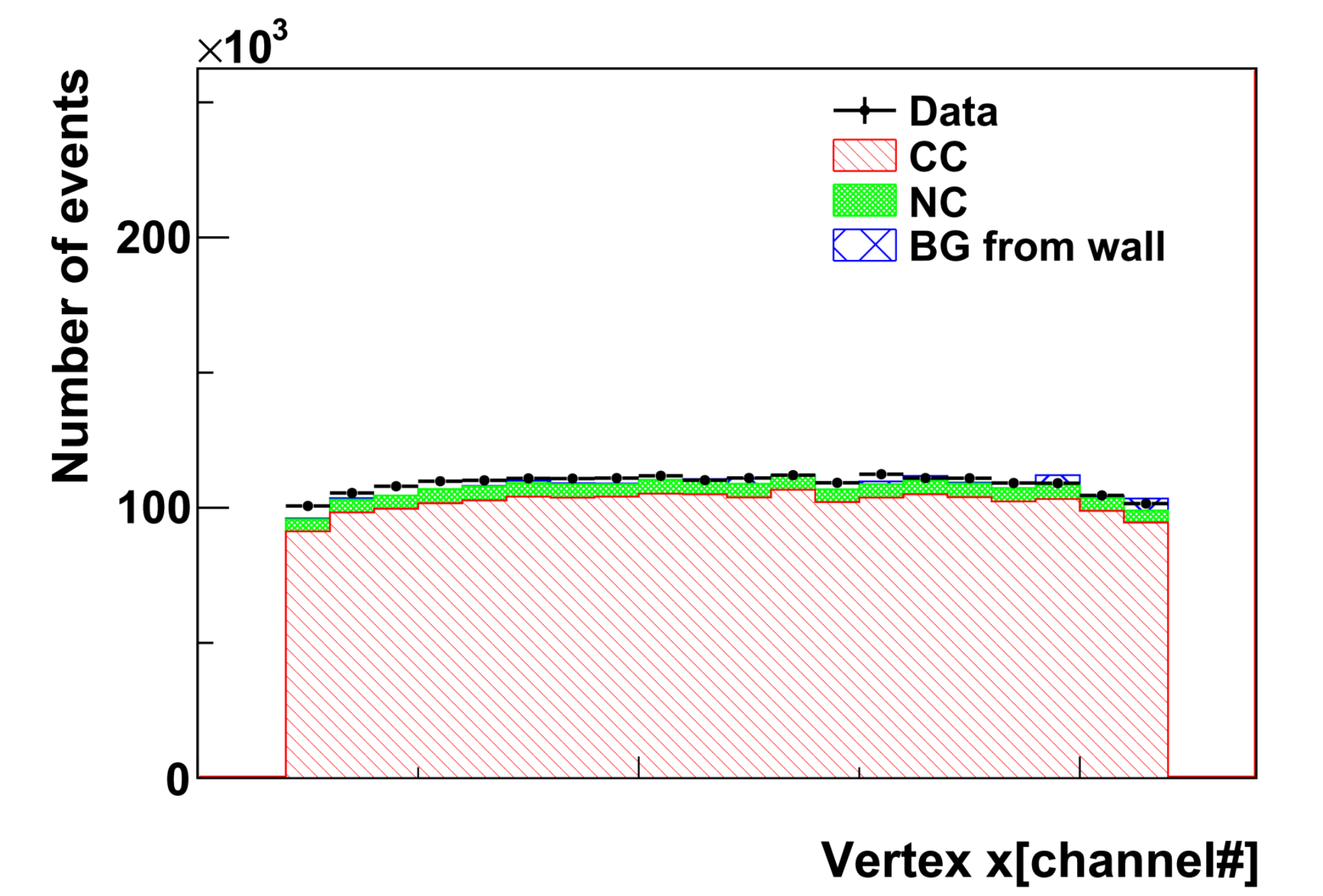}
 \end{center}
\end{minipage}
\begin{minipage}{0.05\hsize}
\end{minipage}
\begin{minipage}{0.475\hsize}
 \begin{center}
  \includegraphics[width=0.9\textwidth]{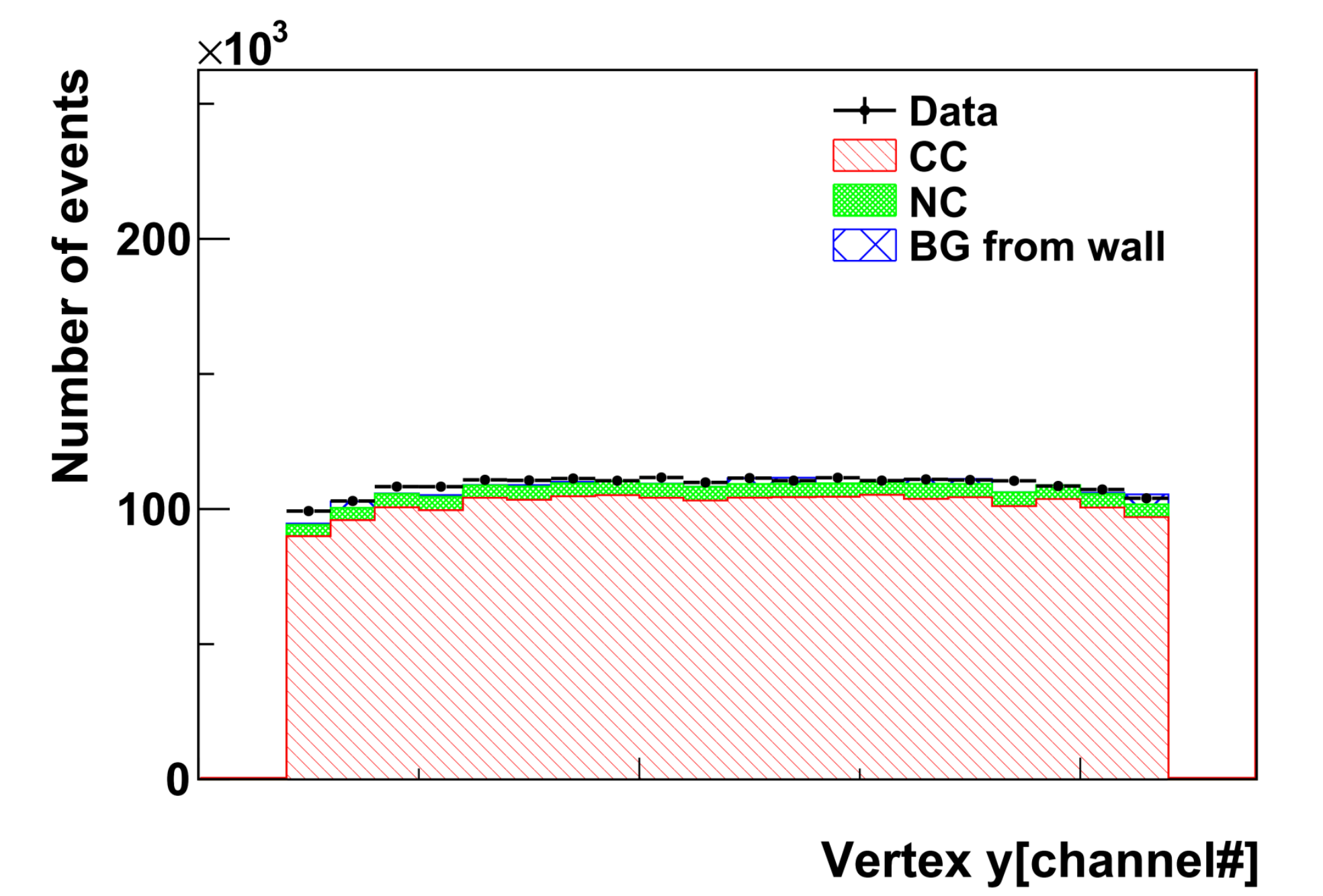}
 \end{center}
 \end{minipage}
 \caption{\modif{Vertex x and y distributions after all the event selections. }}
 \label{fig:verx_very_after}
\end{figure}

\modif{
Figure \ref{fig:verx_very_after} shows the vertex distributions in the x and y direction after all cuts. The fraction of the background in the selected events is 0.4\%. 
}
\subsubsection{Event selection summary}
The result of the event selection is summarized in Table~\ref{tab:selection_summary}. 
The MC simulation includes neutrino interactions in the wall of the experimental hall. 
The number of neutrino interactions in the wall is normalized by the number of beam induced muons in the real data. 
The MC simulation reproduces the reduction in data well. 
\begin{table}[htbp]
\begin{center}
  \caption{Event selection summary for $1.44\times10^{20}$~POT. MC sample is normalized by POT. 
The MC simulation includes neutrino interactions in the wall of the experimental hall. 
The number of neutrino interactions in the wall is normalized by the number of the beam induced muons in the real data. }
  \begin{tabular}{cc||cc|cc}
    \hline
       & selection                        & Data     & (\%)  & MC                & (\%)\\
    \hline
    1  & \# of active planes $>$ 2        & $8.53\times10^6$  &  100  & $9.02\times10^6$  & 100  \\
    2  & PE / active layers $>$ 6.5       & $8.53\times10^6$  &  99.9 & $9.02\times10^6$  & 99.9 \\
    3  & Tracking                         & $8.01\times10^6$  &   94  & $8.40\times10^6$  & 93   \\
    4  & Track matching                   & $7.74\times10^6$  &   91  & $8.10\times10^6$  & 90   \\
    5  & Beam timing                      & $7.73\times10^6$  &   91  & $8.10\times10^6$  & 90   \\
    6  & veto cut                         & $3.30\times10^6$  &   39  & $3.30\times10^6$  & 37   \\
    7  & FV cut                           & $2.18\times10^6$  &   26  & $2.17\times10^6$  & 24   \\
    \hline
  \end{tabular}
  \label{tab:selection_summary}
  \end{center}
\end{table}%
\subsection{Selection efficiency}
\begin{figure}[htbp]
 \begin{center}
  \includegraphics[width=0.7\textwidth]{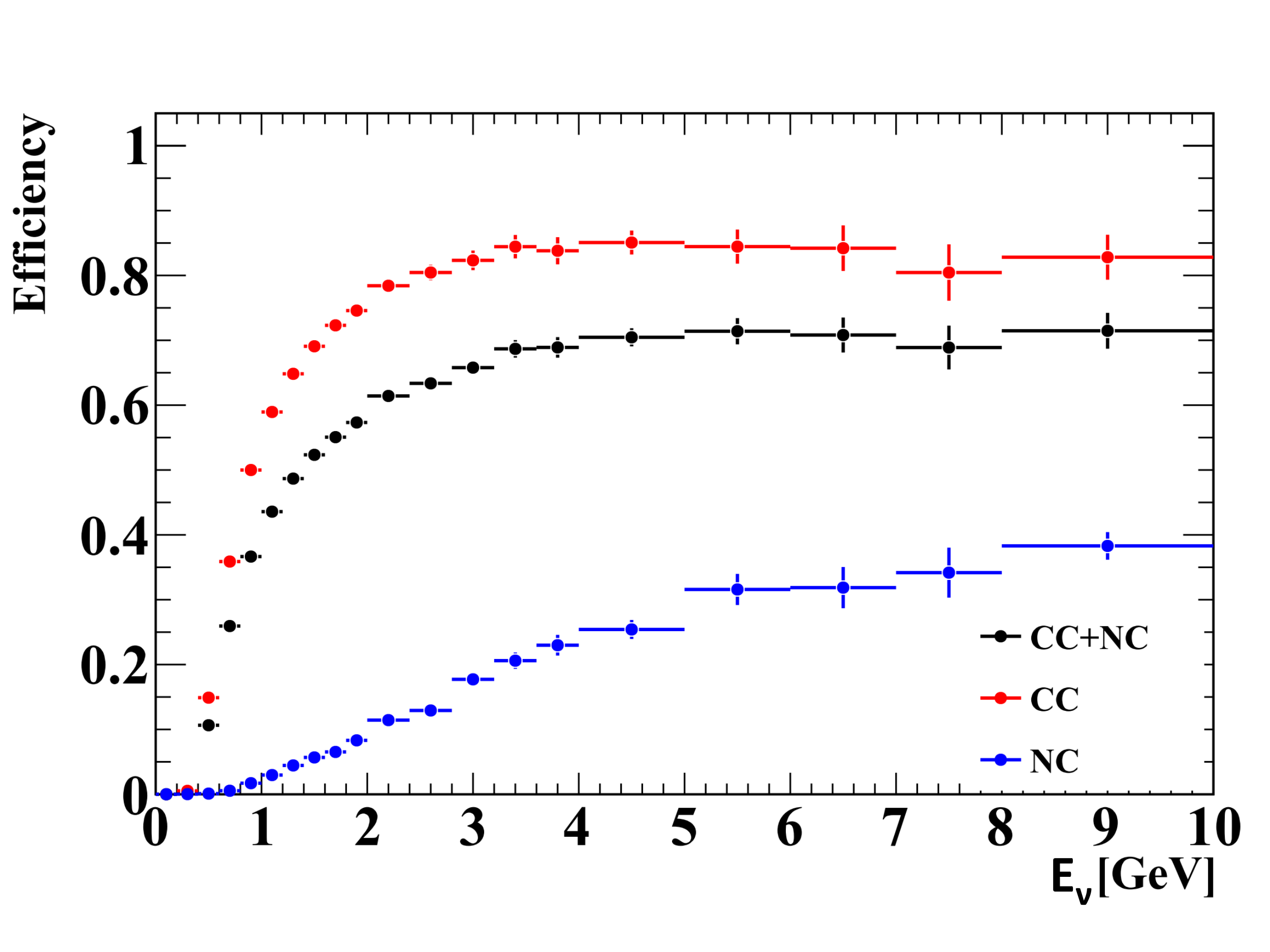}
  \caption{Neutrino event selection efficiency as a function of true neutrino energy.}
  \label{fig:efficiency_curve}
 \end{center}
\end{figure}

The neutrino event selection efficiency as a function of true neutrino energy \kendall{is} estimated by the MC simulation and \kendall{is} shown in Fig.~\ref{fig:efficiency_curve}. 
The $\sim$20\% inefficiency at high energy for the CC interaction is due to the events in which muons are produced with \kendall{a} rather large angle: 
for such events, the muon escapes from the module before it penetrates two iron plates. 
\par
\subsection{Correction factors}\label{subsec:correction}
In order to compare the data with the MC expectation, some corrections are applied to the number of selected neutrino events. 
The correction factors are for the iron target mass and accidental MPPC noise. 
\par
\kendall{The neutrino interactions in the MC simulation are generated with the design mass of iron target plates, so a correction is applied to the neutrino interaction rate in each module individually to based on the difference of the design mass to the measured mass of the module: {$-1\sim+1$\%} for each module. }
\par
Accidental MPPC noise hits sometimes results in mis-identification of the vertex. 
\kendall{The dependence on the event selectionas a function of MPPC noise rate is estimated by a MC simulation where the PE distribution and hit timing distribution are varied to reproduce data noise rates. }
According to this estimation and measured noise rate, the number of selected events is decreased by 3\% with the existence of MPPC noise. 
The number of events in the MC simulation is corrected to account this effect. 
\par
\subsection{Systematic errors}\label{subsec:systematic}
In this section the systematic error on the number of selected events is described. 
We describe both the errors related to the event selection criterion and the correction described in section \ref{subsec:correction}. 
\par
The systematic error of the light yield selection is negligible because the threshold value for the selection (6.5 PE/active layer) is much smaller than the measured light yield \kendall{from} cosmic-ray. 
The systematic error of the tracking efficiency is estimated by comparing the efficiency for several lengths of track between the data and the MC simulation. 
The difference is larger for the shorter track events and the maximum difference of 1.4\% is taken as the systematic error. 
For the selection on the difference of vertex z between x-z and y-z projections, the systematic error is estimated by looking \kendall{at} the change of the selection efficiency when varying the selection threshold. 
The difference (2.7\%) between this and the nominal selection is taken as the systematic error. 
To estimate the systematic error associated with the FV selection, \kendall{the number of neutrino events is re-counted as the fiducial volume is varied. }
In order to check the contamination of the beam induced muon from the edge, the fiducial volume is separated to center, middle and edge parts with the same volume. 
The difference of the event rate per unit volume is 1.1\%, which is taken as the systematic error. 
The discrepancy between the hit efficiency measured with the beam induced muon event and that \kendall{of} the MC simulation is assigned as the uncertainty in the hit detection efficiency. 
The relation between the hit efficiency and the number of selected events is estimated by the MC simulation. 
A systematic error of 1.8\% is assigned. 
The timing measurement is good enough that the systematic uncertainties resulting from the beam timing cut and out-of-beam events are negligible. 
There is a discrepancy between the beam induced muon event rate estimated by the MC simulation and \kendall{that measured from the data.} 
The change \kendall{in} the background contamination fraction from this discrepancy is estimated to be 0.2\%, which is taken as the systematic error for the beam-related background correction. 
\par
The uncertainty of the iron mass measurement, 0.1\%, is taken as the systematic error for the iron mass. 
The change \kendall{in} the selection efficiency due to the time variation of the measured noise rate is estimated to be 0.7\%, which is taken as the systematic error for the MPPC noise correction. 
\par
Table~\ref{tab:syst_error} shows a summary of the systematic errors. 
The total systematic error is calculated as a quadratic sum of the errors and is 3.7\%.
\modif{In addition, there is 2\% uncertainty for the normalization due to the systematic error on the POT measurement. }
\begin{table}[htbp]
\begin{center}
  \caption{Detector systematic errors}
  \begin{tabular}{c|c}
    \hline
     Item                             & Error[\%]     \\ 
    \hline
     Average light yield per active layer & $<$0.1  \\
     Tracking efficiency              & 1.4     \\
     Vertex z matching                & 2.7     \\
     FV cut                           & 1.1     \\
     Hit efficiency                   & 1.8     \\
     out-of-beam events               & $<$0.1  \\
     Beam timing selection            & $<$0.1  \\
     Beam-related background events   & 0.2     \\
     Iron mass                        & 0.1     \\
     Accidental MPPC noise            & 0.7     \\
    \hline
     Total                            & 3.7  \\
    \hline
  \end{tabular}
  \label{tab:syst_error}
  \end{center}
\end{table}%
\section{Results of the measurement}\label{sec:result}
\subsection{Event rate and comparison with the MC expectation}
\par
\begin{figure}[htbp]
 \begin{center}
  \includegraphics[width=1\textwidth]{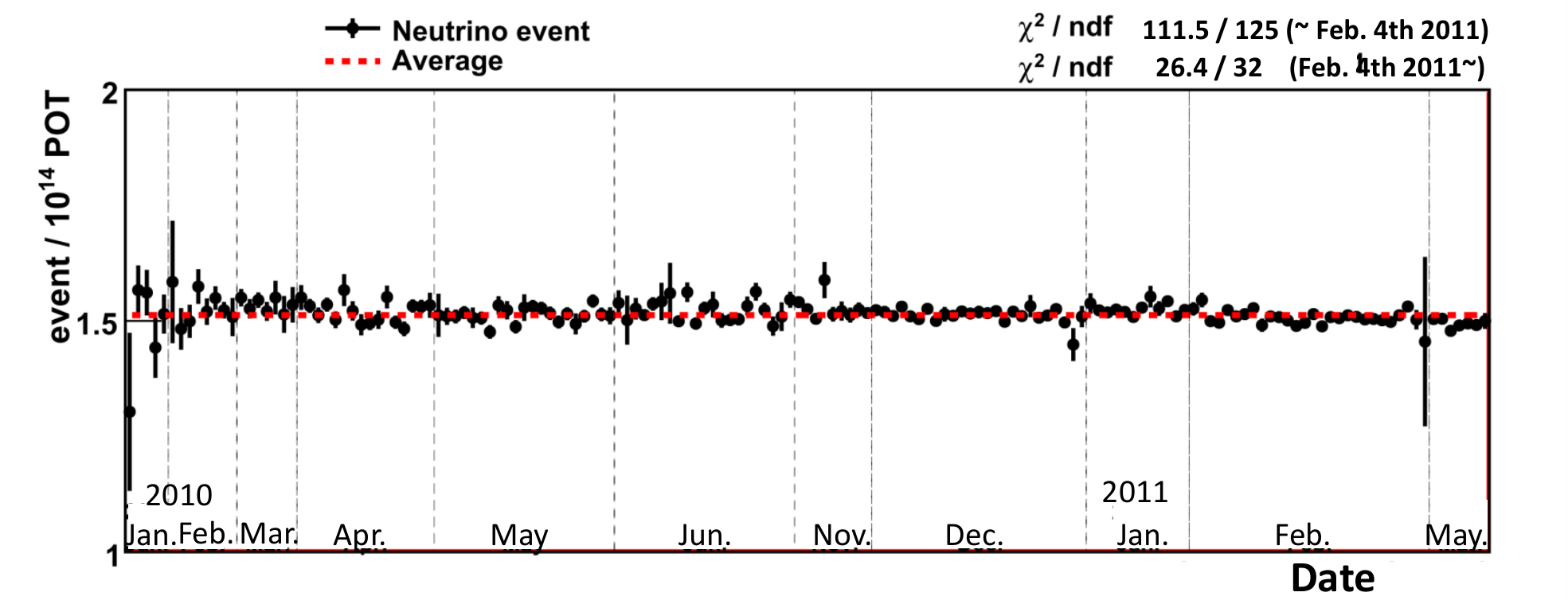}
  \caption{Daily event rate of the neutrino events normalized by protons on target. }
  \label{fig:evtrate_nuevt}
 \end{center}
\end{figure}
\begin{figure}[htbp]
 \begin{center}
  \includegraphics[width=1\textwidth]{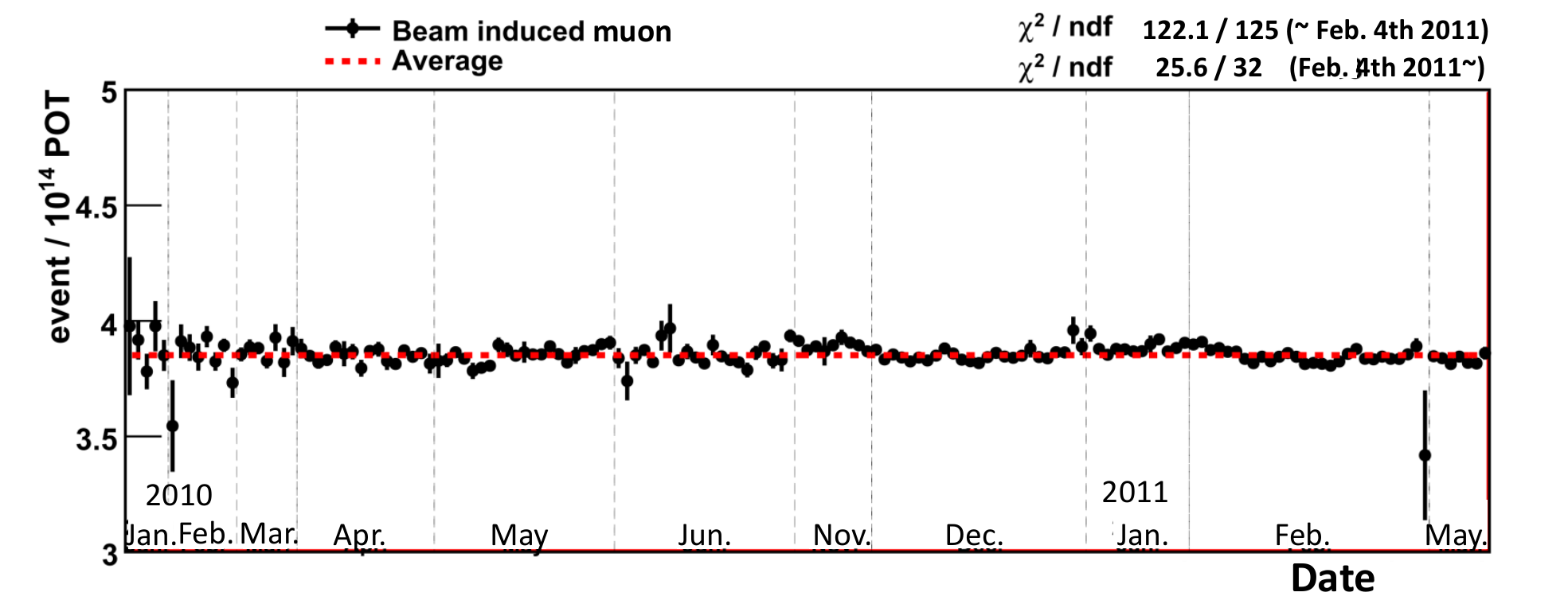}
  \caption{Daily event rate of the beam induced muon events normalized by protons on target. }
  \label{fig:evtrate_dirtmu}
 \end{center}
\end{figure}

Figures~\ref{fig:evtrate_nuevt} and \ref{fig:evtrate_dirtmu} show the daily rates of the neutrino events and the beam induced muon events normalized by protons on target (POT). 
The beam induced muon events are defined as the \kendall{events} rejected by the veto cut or the FV cut in Sec.\ref{sec:analysis}. 
They are measured typically with statistical errors of 1.7\% and 1.1\% each day, respectively. 
\modif{The average event rate is slightly decreased (-0.9\%) on Feb. 4, 2011. 
This is considered to have been caused by a change in the primary beamline condition. 
The muon yield measured by the muon monitor showed a consistent tendency. 
The chi-squared values calculated from the average rates of the neutrino events (beam induced muon events) before and after Feb. 4, 2011 are 111.5 (122.1) for 125 data points and 26.4 (25.6) for 32 data points, respectively. 
The event rate in INGRID before and after Feb. 4, 2011 remained stable within statistical error and the beam intensity is stable during each of these two periods. 
}
\par
The data to MC ratio of the neutrino event rate is calculated to be 1.06$\pm$0.001 (stat.)$\pm$0.04(detector syst.)\modif{$\pm$0.02(POT error)}. 
The uncertainties from the neutrino flux prediction and neutrino interaction model are not included in the systematic error. 
\subsection{Beam direction}
The profile of the beam in the x and y directions \kendall{is} reconstructed with the number of neutrino events in seven horizontal and seven vertical modules, respectively. 
The observed profiles are fitted with a Gaussian function. 
The profile center is defined as the peak of the fit. 
Finally, the beam direction is reconstructed as the direction from the proton beam target position to the measured center at INGRID. 
\par
\begin{figure}[htbp]
  \begin{center}
  \includegraphics[width=1\textwidth]{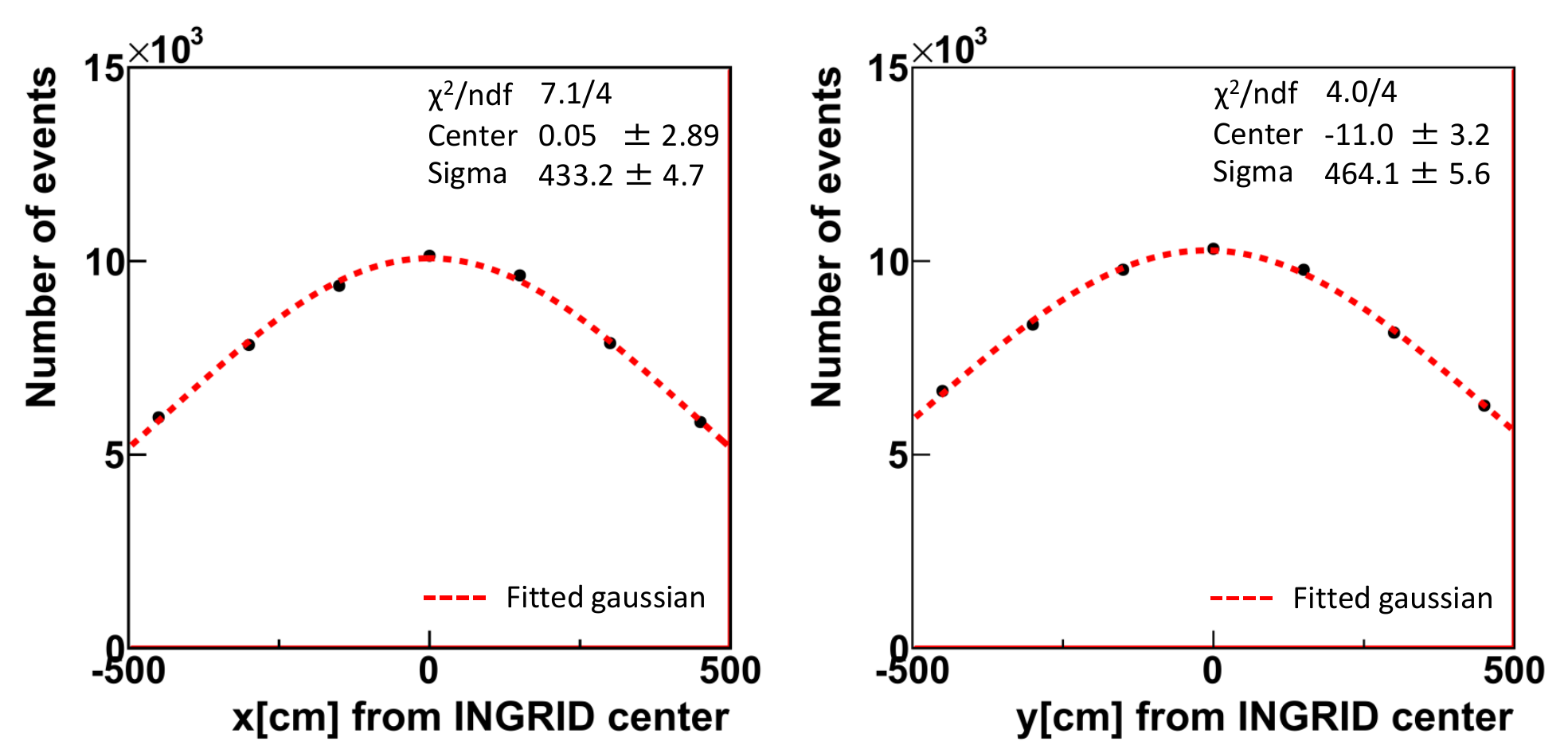}
  \caption{Neutrino beam profiles for x (left) and y (right) directions measured in April 2010.}
  \label{fig:profile_example}
  \end{center}
\end{figure}

In order to monitor the stability of the beam direction, the number of neutrino events is accumulated on a monthly basis. 
Figure~\ref{fig:profile_example} shows the observed profiles in April 2010. 
Black points in this figure show the number of neutrino events in each module and the red dashed lines show the fitted Gaussian.   
\par
\begin{figure}[htbp]
  \begin{center}
  \includegraphics[height=1\textwidth, angle=90]{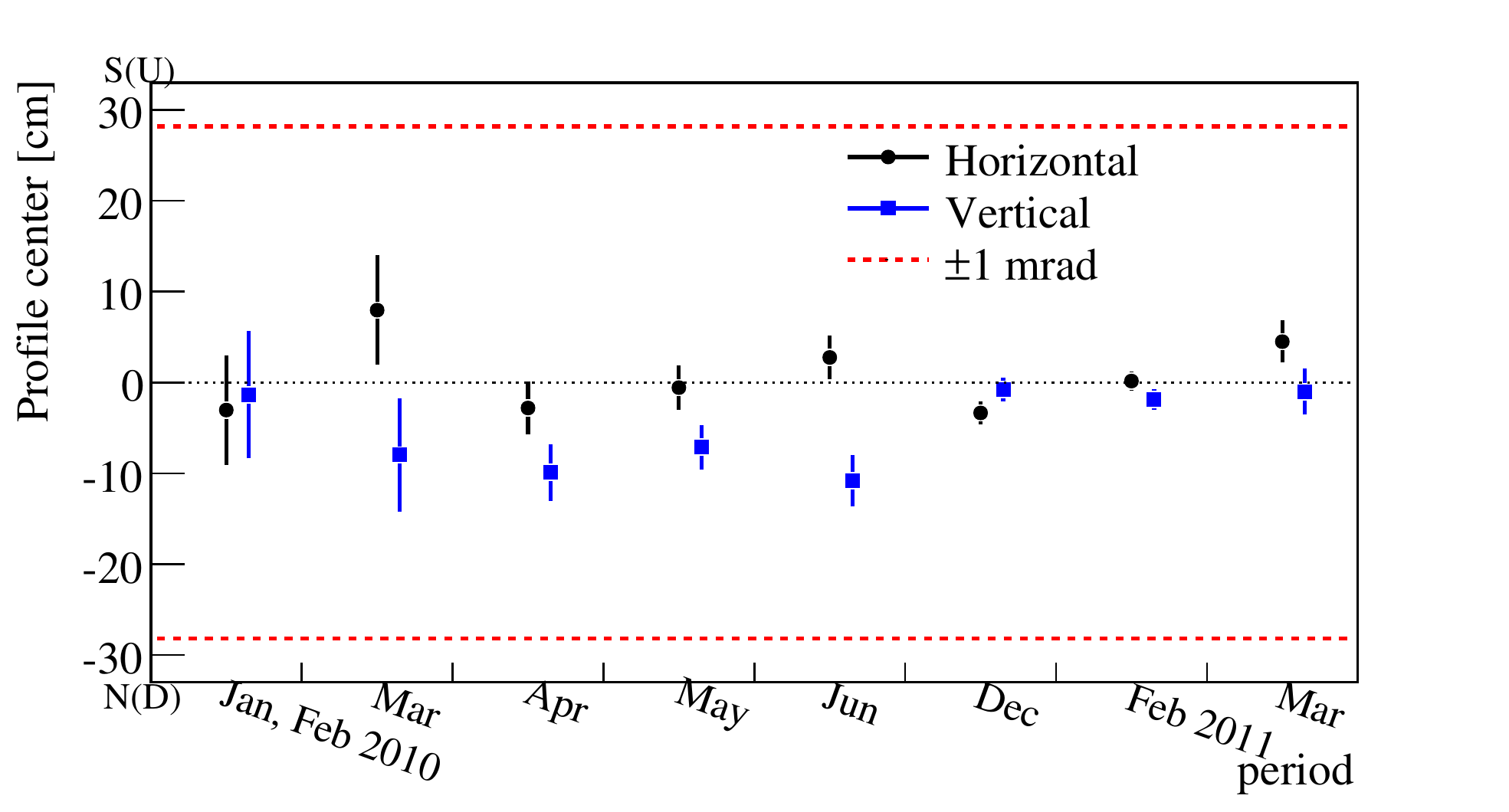}
  \caption{History of the neutrino beam centers.}
  \label{fig:profile}
  \end{center}
\end{figure}

Black and blue points in Fig.~\ref{fig:profile} show the history of the beam centers in the x and y directions. 
All the points were stable well within 28~cm, which corresponds to the requirement of 1 mrad for the beam direction. 
Because beam direction was adjusted in November 2010, the beam centers in the y direction for later months are slightly shifted toward the center. 
\par
The systematic error of the profile center measurement was estimated by a toy MC simulation. 
In the simulation, the number of events at each module is changed within the range of the total detector systematic error of 3.7\%. 
100,000~profiles are generated and RMSs of reconstructed center values are taken as the systematic errors; 9.2~cm and 10.4~cm for the x and y center, respectively. 
\par
From the beam center measurement and the survey between the proton target and the INGRID detectors, the average beam direction in x and y direction are measured as -0.014$\pm$0.025(stat.)$\pm$0.33(syst.) mrad and $-0.107\pm$0.025(stat.)$\pm$0.37 (syst.) mrad, respectively.  
The beam direction is measured with a precision better than the requirement. 
\section{Conclusion}\label{sec:conclusion}
We have reported the muon neutrino beam measurement with the T2K on-axis near neutrino detector, INGRID, for the T2K Run 1 and Run 2 data ($1.44\times10^{20}$~POT in total). 
INGRID consists of 14 identical modules arranged in a cross around the beam center. 
This configuration enables us to sample the beam in a sufficiently wide area to measure the beam center with a minimum \kendall{of} material. 
\par
The neutrino event rate is measured on a daily basis and remains stable within the statistical error, which is typically 1.7\%. 
The data/MC ratio of the event rate is measured to be 1.06$\pm$0.001(stat.)$\pm$0.04(detector syst.)\modif{$\pm$0.02(POT error)}. 
\par
Beam centers in \kendall{the} x and y directions are measured on a monthly basis and are stable well within 28~cm, which corresponds to the required 1~mrad in angle. 
The angle between \kendall{the} expected and measured beam center direction in \kendall{the} x and y directions are -0.014$\pm$ 0.025(stat.)$\pm$0.33(syst.)~mrad and -0.107$\pm$0.025(stat.)$\pm$0.37(syst.)~mrad, respectively.  
\par
We conclude that the neutrino beam intensity and direction remain stable during the measurement based on the precise measurement with INGRID. 
\section*{Acknowledgments}\label{sec:acknowledgements}
It is a pleasure to thank Mr. Taino from Mechanical Support Co. for helping \kendall{with} the construction of INGRID. 
We would like to thank KEK FUJI test beam line group for the test of the INGRID scintillator prototype. 
\par
We thank the Japanese Ministry of Education, Culture, Sports, Science and Technology (MEXT) for their support for T2K. 
The T2K neutrino beamline, the ND280 detector and the Super-Kamiokande detector have been built and operated using funds provided by: the MEXT, Japan; the Natural Sciences and Engineering Research Council of Canada, TRIUMF, the Canada Foundation for Innovation, 
and the National Research Council, Canada; Commissariat \`a l'Energie Atomique, and Centre National de la Recherche Scientifique - Institut National de Physique NuclLeaire et de Physique des Particules, France; 
Deutsche Forschungsgemeinschaft, Germany; Istituto Nazionale di Fisica Nucleare, Italy; 
the Polish Ministry of Science and Higher Education, Poland; the Russian Academy of Sciences, the Russian Foundation for Basic Research, and the Ministry of Education and Science of the Russian Federation, Russia; 
the National Research Foundation, and the Ministry of Education, Science and Technology of Korea, Korea; 
Centro Nacional De F\'isica De Part\'iculas, Astropart\'iculas y Nuclear,
and Ministerio de Ciencia e Innovacion, Spain; the Swiss National Science Foundation and the Swiss State Secretariat for Education and Research, Switzerland; 
the Science and Technology Facilities Council, U.K.; 
and the Department of Energy, U.S.A.
\par
We are also grateful to CERN for their generous donation of the UA1/NOMAD magnet for our use and their assistance with their generous donation of the HERA-B magnet movers. 
\par
In addition, the participation of individual researchers and institutions in the construction of the T2K experiment has been further supported by funds from: 
the European Research Council; 
the Japan Society for the Promotion of Science Fellowship for Foreign Researchers program; 
the U.S. Department of Energy Outstanding Junior Investigator (OJI) Program and Early Career program; 
and the A. P. Sloan Foundation. 
\par
The authors also wish to acknowledge the tremendous support provided by the collaborating institutions, including but not limited to: 
Ecole Polytechnique - Palaiseau; and the State University of New York at Stony Brook, Office of the Vice President for Research.






\bibliographystyle{model1a-num-names}
\bibliography{<your-bib-database>}



\end{document}
\else

\fi
